\pdfoutput=1
\documentclass[prd,byrevtex
,preprint
,nofootinbib,
,tightenlines
,showkeys
,showpacs]{revtex4}
\usepackage{amsmath,amsxtra,amssymb}
\usepackage{graphicx}
\usepackage{float}
\newcommand{\bs}[1]{\boldsymbol{#1}}
\newcommand{\pa}{\partial}

\newcommand{\del}{\delta}
\newcommand{\ep}{\epsilon}

\begin{document}
\title{Entanglement Structure in Expanding Universes}
\author{Yasusada Nambu}
\email{nambu@gravity.phys.nagoya-u.ac.jp}
\affiliation{Department of Physics, Graduate School of Science, Nagoya 
University, Chikusa, Nagoya 464-8602, Japan}

\begin{abstract}
 We investigate entanglement of a quantum field in de Sitter
  spacetime using a particle detector model. By considering the
  entanglement between two comoving detectors interacting with a
  scalar field, it is possible to detect the entanglement of the
  scalar field by swapping it to detectors. For the massless minimal
  scalar field, we find that the entanglement between the detectors
  cannot be detected when their physical separation exceeds the Hubble
  horizon scale. This behavior supports the appearance of the
  classical nature of quantum fluctuations generated during the
  inflationary era.
\end{abstract}
\pacs{04.62.+v, 03.65Ud}
\keywords{entanglement; particle detector; inflation; measurement}
\maketitle


\vspace{-12pt}
\section{Introduction}

According to the inflationary scenario of cosmology, all structure in the Universe can be traced back to primordial quantum fluctuations during an accelerated expanding phase of the very early universe. Short wavelength quantum fluctuations generated during inflation are considered to lose quantum nature when their wavelengths exceed the Hubble horizon length. Then, the statistical property of generated fluctuations can be represented by classical distribution functions. This is the assumption of the quantum to classical transition of quantum fluctuations generated by the inflation. As the structure in the present Universe is classical objects, we must explain or understand how this transition occurred and how the quantum fluctuations changed to classical fluctuations~\cite{KieferC:AL2:2009}. When we calculate a correlation function of observables between two spatially separated regions, we have a possibility that the quantum correlation function cannot be reproduced using a local classical probability distribution function if these two regions are entangled~\cite{GenoveseM:ASL2:2009, BallJL:PLA359:2006, FuentesI:PRD83:2010} and the classical locality is violated. In other words, we cannot regard the quantum fluctuations as the classical stochastic fluctuations as long as the system is entangled. Therefore, it is important to clarify the relation between the entanglement and the appearance of the classical nature to fully understand the mechanism of the quantum to classical transition of primordial fluctuations.

We have investigated the problem of quantum to classical transition from the viewpoint of entanglement~\cite{NambuY:PRD78:2008,NambuY:PRD80:2009,NambuY:PRD84:2011}. In our previous study, we considered the intrinsic entanglement of quantum fields~\cite{NambuY:PRD78:2008,NambuY:PRD80:2009}. We defined two spatially separated regions in the inflationary universe and investigated the bipartite entanglement between these regions. It was found that the entanglement between these two regions becomes zero after their physical separation exceeds the Hubble horizon. This behavior of the bipartite entanglement confirms our expectation that the long wavelength quantum fluctuations during inflation behave as classical fluctuations and can become seed fluctuations for the structure formation in the Universe. These analysis concerning the entanglement of quantum fluctuations in the inflationary universe relies on the separability criterion for continuous bipartite systems~\cite{SimonR:PRL84:2000,DuanL:PRL84:2000} of which dynamical variables are continuous. The applicability of this criterion is limited to systems with Gaussian states: the wave function or the density matrix of the system is represented in a form of Gaussian functions. Thus, we cannot say anything about the entanglement for the system with non-Gaussian state such as excited states and thermal states. Furthermore, from a viewpoint of observation or measurement, information on quantum fluctuations can be extracted via interaction between quantum fields and measurement devices. Hence, it is more natural to consider a setup that the entanglement of quantum field is probed using particle detectors~\cite{UnruhWG:PRD14:1976,BirrellND:CUP:1982}.

In this direction, we considered the detection of the entanglement of scalar fields using particle detectors with two internal energy levels interacting with fields~\cite{NambuY:PRD84:2011}. By preparing two spatially separated equivalent detectors interacting with the scalar field, we can extract the information on entanglement of the scalar field by evaluating the entanglement between these two detectors. Since a pair of such detectors is a two-qubit system, we have the necessary and sufficient condition for entanglement of this system~\cite{PeresA:PRL77:1996,HorodeckiM:PLA223:18}. Using this setup, B. Reznik~\textit{el al.}~\cite{ReznikB:FP33:2003,ReznikB:PRA71:2005} studied the entanglement of the Minkowski vacuum. They showed that an initially non-entangled pair of detectors evolved to an entangled state through interaction with the scalar field. Because the entanglement cannot be created by local operations, this implies that the entanglement of the quantum field is transferred to a pair of detectors. M.~Cliche and A.~Kempf~\cite{ClicheM:PRA81:2010} constructed the information-theoretic quantum channel using this setup and evaluated the classical and quantum channel capacities as a function of the spacetime separation. G.~V.~Steeg and N.~C.~Menicucci~\cite{SteegGV:PRD79:2009,MartinMartinezE:CQG29:2012}
investigated the entanglement between detectors in de Sitter spacetime and they concluded that the conformal vacuum state of the massless scalar field can be discriminated from the thermal state using the measurement of entanglement. Recently, a protocol to extract past-future vacuum entanglemnt is proposed~\cite{SabinC:PRL109:2012}. In our paper~\cite{NambuY:PRD84:2011}, it was found that the entanglement between the detectors becomes zero after their physical separation exceeds the Hubble horizon. Furthermore, the quantum discord, which is defined as the quantum part of total correlation, approaches zero on the super-horizon scale. These behaviors support the appearance of the classical nature of the quantum fluctuation generated during the inflationary era.

In this article, we present our study on the entanglement structure of the quantum field in the expanding universe using the particle detector model. The main purpose is to introduce the detail of our investigation on this subject. This paper is organized as follows. In Section~2, we present our setup of the detector system. Then, in Section~3, we summarize the Wightman functions of the scalar field. In Section~4, we evaluate entanglement and correlations for quantum fields in de Sitter spacetime using asymptotic approximation. In Section~5, we present our result of the negativity obtained using numerical calculation. The result covers almost all parameter range of the model. In Section~6, we discuss the relation between the structure of detector and the observable strength of spatial correlation. Section~7 is devoted to summary. We use units in which $c=\hbar=G=1$ throughout the paper.

\section{ Detector Model}

In this section, we explain the detail of the particle detector model~\cite{UnruhWG:PRD14:1976,BirrellND:CUP:1982} that we use to measure the entanglement of the quantum field. We consider a detector interacting with a scalar field $\phi$. The detector has two energy level states $|\uparrow\rangle, |\downarrow\rangle$ with energy difference $\Omega$. The Hamiltonian of the detector is
\begin{equation}
H_0=\frac{\Omega}{2}\Bigl(|\uparrow\rangle\langle\uparrow|
-|\downarrow\rangle\langle\downarrow|\Bigr)
\end{equation}
Thus our detector model is a qubit system. The interaction Hamiltonian is
\begin{equation}
V=g(t)(\sigma^++\sigma^-)\phi(\bs{x}(t))
\end{equation}
where $\bs{x}(t)$ is the world line (location) of the detector and $\sigma^+, \sigma^-$ are raising and lowering operators for the detector's state: $\sigma^+=|\uparrow\rangle\langle\downarrow|,
\sigma^-=|\downarrow\rangle\langle\uparrow|$. A function $g(t)$ represents the strength of the coupling between the detector and the scalar field. The total Hamiltonian of the system is
\begin{equation}
H=H_0+V+H_\phi
\end{equation}
where $H_\phi$ is the Hamiltonian for the scalar field. Under the Schr\"{o}dinger representation, the state of the total system composed of the detector and the scalar field obeys
\begin{equation}
i\frac{\pa}{\pa t}|\Psi(t)\rangle=(H_0+V+H_\phi)|\Psi(t)\rangle
\end{equation}
By introducing the evolution operator for the free part of the total Hamiltonian as
\begin{equation}
i\frac{\pa}{\pa t}U_0(t,t_0)=(H_0+H_\phi)U_0(t,t_0),\qquad U_0(t_0,t_0)=1
\end{equation}
we can define the interaction representation of the state $|\tilde\Psi(t)\rangle=U_0^\dag(t)|\Psi(t)\rangle$. This state obeys
\begin{equation}
i\frac{\pa}{\pa t}|\tilde\Psi(t)\rangle=U_0^\dag VU_0|\tilde\Psi(t)\rangle=\tilde V|\tilde\Psi(t)\rangle,\qquad
\tilde V=g(e^{i\Omega t}\sigma^++e^{-i\Omega t}\sigma^{-})\phi(t,\bs{x}(t))
\end{equation}
The solution of this equation is 
\begin{equation}
|\tilde\Psi(t)\rangle
=\left(1-i\int_{t_0}^tdt_1\tilde V(t_1)-\frac{1}{2}\int_{t_0}^tdt_1
\int_{t_0}^tdt_2T[\tilde V(t_1)\tilde V(t_2)]+\cdots\right)
|\tilde\Psi(t_0)\rangle \label{eq:final-wavefunc}
\end{equation}
where the time ordering is defined by 
$$
T[A(t_1)A(t_2)]=
\begin{cases}
A(t_1)A(t_2),\quad & t_1>t_2 \\
A(t_2)A(t_1),\quad & t_2>t_1
\end{cases}
$$
Let us introduce the following operators:
\begin{align}
&\Phi^{\pm}=\int_{t_0}^t dt_1g(t_1)e^{\pm i\Omega t_1}\phi(t_1,\bs{x}(t_1)), \\
&S=-i\int_{t_0}^tdt_1\tilde V(t_1)=-i\sum_{j=\pm}\Phi^j(t)\sigma^j
\end{align}
To detect vacuum fluctuation of the scalar field, we prepare the
initial state $|\Psi_0\rangle=|\downarrow\rangle|0\rangle$ \linebreak
where $|0\rangle=\prod_{\bs{k}}|0_{\bs{k}}\rangle$ is the vacuum state
of the scalar field. Up to the leading order of perturbation, only the one
particle state of the scalar field contributes and
\begin{align*}
&\Phi^+|0\rangle=|1\rangle\langle 1|\Phi^+|0\rangle,\\
&\Phi^{-}\Phi^{+}|0\rangle=\Phi^-|1\rangle\langle 1|\Phi^+|0\rangle
=|0\rangle\langle 0|\Phi^{-}|1\rangle\langle1|\Phi^+|0\rangle+\text{(two particle state)}.
\end{align*}
Then using Equation~\eqref{eq:final-wavefunc},
the final state of the total system up to the lowest order of the
perturbation including the interaction between the detector and the
scalar field is 
\begin{align}
|\tilde\Psi_\text{f}\rangle &=
\left(1+S+\frac{1}{2}T[SS]\right)|\tilde\Psi_0\rangle
\notag \\
&=\left(1-\frac{1}{2}\langle 0|T[\Phi^-\Phi^+]|0\rangle\right)|\downarrow\rangle|0\rangle
-i\langle 1|\Phi^+|0\rangle\,|\uparrow\rangle|1\rangle
\end{align}
The total system can be represented by the pure state density operator $\rho_T=|\tilde\Psi\rangle\langle\tilde\Psi|$. By tracing over the degrees of freedom of the scalar field, the density matrix for the detector becomes
\begin{align}
\rho&\equiv\mathrm{tr}_\phi\rho_T=\sum_n\langle n|\tilde\Psi_\text{f}\rangle
\langle\tilde\Psi_\text{f}|n\rangle \notag \\
&=(1-E)
|\downarrow\rangle\langle\downarrow|
+E|\uparrow\rangle\langle\uparrow| =\begin{pmatrix} E & 0 \\ 0 & 1-E \end{pmatrix}
\end{align}
where the basis $\{|\uparrow\rangle, |\downarrow\rangle\}$ is used for the matrix representation of the state. We have introduced the response function~\cite{BirrellND:CUP:1982}
\begin{equation}
\label{eq:E}
E\equiv \int_{-\infty}^\infty dt_1\int_{-\infty}^\infty dt_2g_1g_2 e^{-i\Omega(t_1-t_2)}\langle 0|\phi_1\phi_2|0\rangle
\end{equation}
This quantity represents the transition probability from the down state to the up state of the detector.

To measure quantum correlations of the scalar field, we consider two equivalent detectors interacting with the scalar field. In this case, the interaction Hamiltonian is
\begin{equation}
\tilde V=g(e^{i\Omega t}\sigma^+_A+e^{-i\Omega t}\sigma^-_A)\phi(t,\bs{x}_A(t))+g(e^{i\Omega t}\sigma^+_B+e^{-i\Omega t}\sigma^-_B)\phi(t,\bs{x}_B(t))
\end{equation}
where $\bs{x}_{A,B}(t)$ represent the world lines of each detector. In the same way as the single detector case, we introduce the following operators:
\begin{align*}
&S=-i\int_{t_0}^t dt_1\tilde V(t_1)=-i\sum_{j=\pm}\sum_{k=A,B}\Phi^{j}_k(t)\sigma^j_k\\
&\Phi^{\pm}_k=\int_{t_0}^t dt_1 g(t_1)e^{\pm i\Omega t_1}\phi(t_1,\bs{x}_k(t_1)),\quad k=A,B
\end{align*}
We prepare the initial state $|\Psi_0\rangle=|\downarrow\downarrow\rangle|0\rangle$ to detect the entanglement of the vacuum state of the scalar field. Then using Equation~\eqref{eq:final-wavefunc}, the final state of the total system up to the lowest order of the perturbation including the interaction between the detectors and the scalar field is
\begin{align}
|\tilde\Psi_\text{f}\rangle&
=\left(1+S+\frac{1}{2}T[SS]\right)|\tilde\Psi_0\rangle
\notag \\
&=\Bigl(|\downarrow\downarrow\rangle+d_3|\uparrow\uparrow\rangle
+d_4|\downarrow\downarrow\rangle\Bigr)|0\rangle
+\Bigl(d_1|\uparrow\downarrow\rangle+d_2|\downarrow\uparrow\rangle\Bigr)
|1\rangle \label{eq:final-wavefunc2}
\end{align}
where the coefficients are defined by (Figure~1)
\begin{align*}
&d_1=-i\langle 1|\Phi_A^+|0\rangle,\quad d_2=-i\langle 1|\Phi_B^+|0\rangle,
\\
&d_3=-\langle 0|T[\Phi_A^+\Phi_B^+]|0\rangle,\quad d_4=-\frac{1}{2}\Bigl(
\langle 0|T[\Phi_A^-\Phi_A^+]|0\rangle+\langle 0|T[\Phi_B^-\Phi_B^+]| 0\rangle\Bigr)
\end{align*}
\begin{figure}[H]
\centering
\includegraphics[width=0.9\linewidth,clip]{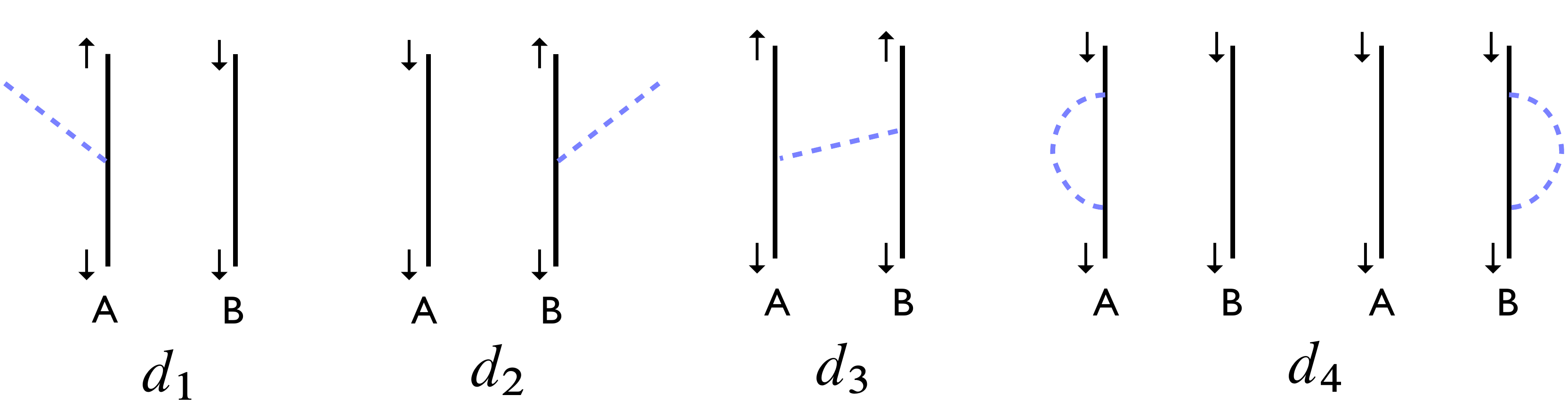}
\caption{Emission and absorption processes contribute to the final state \eqref{eq:final-wavefunc2}. These diagrams represent virtual processes and do not conserve energy.}
\end{figure}
\noindent By tracing over the scalar field, the density matrix for the detectors system is given by
\begin{align}
\rho_{AB}&=\sum_{n}\langle n|\tilde\Psi\rangle\langle\tilde\Psi|n\rangle \notag\\
&=
\begin{pmatrix}
0 & 0 & 0& d_3 \\ 0 & d_1d_1^* & d_1d_2^* & 0 \\
0 & d_1^*d_2 & d_2d_2^* & 0 \\
d_3^* & 0 & 0 & 1+d_4+d_4^*
\end{pmatrix}
\equiv\begin{pmatrix}
0 & 0 & 0 & X \\
0 & E & E_{AB} & 0 \\
0 & E_{AB} & E & 0 \\
X^* & 0 & 0 & 1-2E
\end{pmatrix}
\label{eq:bi-state}
\end{align}
where we use the basis $|AB\rangle=\{|\uparrow\uparrow\rangle,|\uparrow\downarrow\rangle,
|\downarrow\uparrow\rangle, |\downarrow\downarrow\rangle\}$.
The components of the density matrix are given by
\begin{align}
&E= d_1d_1^*=\langle0|\Phi^-_A\Phi^+_A| 0\rangle=\int dt_1dt_2g_1g_2 e^{-i\Omega(t_1-t_2)}
\langle\phi_1\phi_2\rangle=d_2d_2^* \notag\\
&E_{AB}=d_1d_2^*=\langle0|\Phi^-_A\Phi^+_B| 0\rangle
=\int dt_1dt_2g_1g_2 e^{-i\Omega(t_1-t_2)}\langle
\phi_{1A}\phi_{2B}\rangle \label{eq:XE}\\
&X=d_3=-2\int_{t_1>t_2}dt_1dt_2g_1g_2 e^{i\Omega(t_1+t_2)}\langle\phi_{1A}\phi_{2B}\rangle \notag
\end{align}

For the purpose of detecting entanglement of quantum fields, we consider the negativity of the system. The negativity is defined using the eigenvalues of the partially transposed density matrix. For the bipartite state $\rho_{AB}$, let us consider the partial transpose of the state with respect to B. Then, assuming that the dimension of the system is $2\times 2$ or $3\times 3$, the following theorem holds for the partially transposed state $\rho'_{AB}$~\cite{HorodeckiP:PLA232:1997}:
$$
\text{All eigenvalues of }\rho'\text{ are positive}\quad\Leftrightarrow\quad\text{The bipartite state is separable}
$$
For the state \eqref{eq:bi-state}, eigenvalues of the partially transposed state are $0, 1-2E, E\pm|X|$. Thus, for the state~satisfying
$$
|X|\ge E $$
the theorem says that the bipartite system is entangled. As a measure of the entanglement between two detectors, we adopt the negativity~\cite{VidalG:PRA65:2002}. The negativity is defined using the eigenvalues of the partially transposed density matrix. In the present case, the negativity is
\begin{equation}
\mathcal{N}=\mathrm{max}\left[0,|X|-E\right]
\end{equation}
In this paper, we designate the following quantity as the negativity
\begin{equation}
\label{eq:4}
\mathcal{N}=|X|-E
\end{equation}
The negativity gives the necessary and the sufficient condition of the entanglement for two-qubit systems~\cite{HorodeckiP:PLA232:1997}. Two detectors are entangled when $\mathcal{N}>0$ and separable when $\mathcal{N}<0$. For separable initial states of detectors, $\mathcal{N}>0$ after interaction with the scalar field implies the scalar field is entangled because entanglement cannot be generated by local operations provided that two detectors are spatially~separated.

\section{Scalar Fields and Wightman Functions}

We consider the scalar field in de Sitter spacetime with a spatially flat slice. The metric is
$$
ds^2=-dt^2+e^{2Ht}d\bs{x}^2=a^2(\eta)(-d\eta^2+d\bs{x}^2),\qquad
\eta=-\frac{e^{-Ht}}{H}
$$
where $\eta$ is the conformal time and $H$ is the Hubble constant. The equation of motion of the scalar field~is
\begin{equation}
\label{eq:2}
\ddot\phi+3H\dot\phi-\frac{1}{a^2}\nabla^2\phi+(m^2+\xi R(t))\phi=0
\end{equation}
A constant $\xi$ represents the coupling between the scalar field and the spacetime curvature. The quantized field is represented as
$$
\phi(t,\bs{x})=\int\frac{d^3k}{(2\pi)^{3/2}}\left(f_k(t)\hat a_{\bs{k}}+f_k^*\hat a^\dag{}_{\!\!\!-\bs{k}}\right)e^{i\bs{k}\cdot\bs{x}}
$$
where $[\hat a_{\bs{k}_1}, \hat a_{\bs{k}_2}^\dag]=\del^3(\bs{k}_1-\bs{k}_2), [\hat a_{\bs{k}_1}, \hat a_{\bs{k}_2}]=[\hat a_{\bs{k}_1}^\dag, \hat a_{\bs{k}_2}^\dag]=0$ and the vacuum state is defined by $\hat a_{\bs{k}}|0\rangle=0$. The mode function $f_k(t)$ obeys
\begin{equation}
\label{eq:1}
\ddot f_k+3H\dot f_k+\frac{k^2}{a^2}f_k+\left(m^2+\xi R\right)f_k=0
\end{equation}
with the normalization $f_k\dot f_k^*-f_k^*\dot f_k=i/a^3$. The Wightman function is defined by
\begin{align}
D^+(t_1,\bs{x}_1;t_2,\bs{x}_2)&=\langle\phi(t_1,\bs{x}_1)\phi(t_2,\bs{x}_2)
\rangle \label{eq:wight2}\\
&=\frac{1}{2\pi^2}\int_0^\infty dkk^2j_0(kr)f_k(t_1)f_k^*(t_2),\quad r=|\bs{x}_1-\bs{x}_2| \notag
\end{align}
Using the rescaled field variable $\varphi=a\phi$ and the conformal time, the equation of motion of the massless scalar field is
\begin{equation}
\varphi''-\nabla^2\varphi+(6\xi-1)\frac{a''}{a}\varphi=0,\quad
'=\frac{\pa}{\pa\eta}
\end{equation}
For $\xi=1/6$, the field $\varphi$
obeys the same equation as the massless scalar field in the Minkowski spacetime (the conformal invariant scalar field) and $\xi=0$
corresponds to the minimally coupled scalar field. We assume that the detectors are comoving with respect to the cosmic expansion and the physical distance between them increases with time proportional to the scale factor.

\subsection{Minkowski Vacuum and Thermal State}

In Minkowski spacetime, the Wightman functions for the vacuum state and the thermal state with temperature $T$ are
\begin{align}
&D^+_M=\frac{-1}{4\pi^2}\frac{1}{(\Delta t-i\ep)^2-r^2},\quad
\Delta t=t_1-t_2\\
&D^+_T=\frac{T}{8\pi r}\left[\coth\pi T(r-\Delta t+i\ep)+\coth\pi T(r+\Delta t-i\ep)\right]
\end{align}
where we introduced a small parameter $\ep>0$ to regularize ultraviolet divergence of the $k$ integral \eqref{eq:wight2}. By introducing new time variables
$$
x=\frac{t_1+t_2}{2},\quad y=\frac{t_1-t_2}{2}
$$
they become
\begin{align}
&D_M^{+}=-\frac{1}{8\pi^2}\frac{1}{(y-i\ep)^2-(r/2)^2}\\
&D_T^{+}=\frac{T}{8\pi r}\left[\coth\pi T(r-2y+i\ep)+\coth\pi T(r+2y-i\ep)\right]
\end{align}
\subsection{Conformal Vacuum}

For the conformal massless scalar field $\xi=1/6$, the Wightman function is
\begin{align}
D_C^+(x_1,x_2)&=-\frac{H^2}{4\pi^2}\left[4\sinh^2\left(\frac{H}{2}(\Delta t-i\ep)\right)
-e^{H(t_1+t_2)}r^2\right]^{-1} \notag \\
&=\frac{H^2}{16\pi^2}\left[-\sinh^2(H(y-i\ep))+e^{2Hx}(Hr/2)^2\right]^{-1}
\end{align}
\subsection{Minimal Scalar}

For the minimal scalar field, we present the detail of the derivation of the Wightman function. Assuming the Bunch--Davies vacuum state, the mode function of the minimal massless scalar field in de Sitter spacetime is
\begin{equation}
\phi_k=-\frac{H}{\sqrt{2k}}\left(\eta-\frac{i}{k}\right)e^{-ik\eta}
\end{equation}
The Wightman function is
\begin{align*}
& D^+_{\text{dS}}(t_1,\bs{x}_1,t_2,\bs{x}_2)\\
\quad &=\frac{H^2\eta_1\eta_2}{4\pi^2 r}\int_0^\infty dk\sin kr\,e^{-\ep k}e^{-ik\Delta\eta}
+\frac{H^2}{4\pi^2 r}\int_{k_0}^\infty dk\sin kr\left(-\frac{\pa}{\pa k}\right)
\left(\frac{e^{-ik\Delta\eta-k\ep}}{k}\right)\\
\quad &\equiv D_1+D_2,\quad\Delta\eta=\eta_1-\eta_2
\end{align*}
where we introduced a small $\ep>0$ to regularize the ultraviolet divergence of the $k$ integral and a small positive constant $k_0\ll H$ to regularize infrared divergence of the $k$ integral. Physically, $k_0^{-1}$ corresponds to the horizon scale at the onset of the inflation. $D_1$ and $D_2$ are given by {$$
\mathrm{Ei}(-x)=-\int_x^\infty\frac{dt}{t}e^{-t} $$}
\begin{align}
&D_1(x,y,r)
=\frac{H^2}{16\pi^2}\left[-\sinh^2(H(y-i\ep))+e^{2Hx}(Hr/2)^2\right]^{-1}\\
&D_2(x,y,r)=-\frac{H^2}{8\pi^2}\Biggl\{\mathrm{Ei}\left[-\frac{ik_0}{H}
\left(-Hr+2e^{-Hx}\sinh(H(y-i\ep))
\right)\right]\\
&\qquad\qquad
+\mathrm{Ei}\left[-\frac{ik_0}{H}
\left(Hr+2e^{-Hx}\sinh(H(y-i\ep))\right)
\right]\Biggr\}+\frac{H^2}{4\pi^2} \notag
\end{align}
$D_1$ is the same as the Wightman function for the conformal invariant massless scalar field.

An equal time correlation function $\langle\phi(t,\bs{x}_1)\phi(t,\bs{x}_2)\rangle$ is obtained by taking $y=0$:
\begin{align*}
&D_1(t,r)=\frac{H^2}{4\pi^2}\frac{e^{-2Ht}}{(Hr)^2}=\frac{1}{4\pi^2}
\frac{1}{r_p^2}, \quad r_p=re^{Ht}\\
&D_2(t,r)=-\frac{H^2}{8\pi^2}\left[\mathrm{Ei}(ik_0r)+\mathrm{Ei}(-ik_0r)
\right]+\frac{H^2}{4\pi^2}
\end{align*}
For $k_0r \ll 1$,
$$
D_2\approx -\frac{H^2}{4\pi^2}\left[\ln(k_0r)+\gamma-1\right]
=-\frac{H^2}{4\pi^2}\left[\ln(Hr_p)-N+\gamma-1\right] $$
where we have introduced the onset time $t_0$ of the inflation by
\begin{equation}
k_0/H=e^{Ht_0}
\end{equation}
The e-foldings from $t_0$ to $t$ is $N=H(t-t_0)$.


As we can observe in Figure~2, the correlation on the sub-horizon scale is $\propto 1/r_p^2$ and this behavior is the same as the Minkowski vacuum and the conformal vacuum. At the Hubble horizon scale $r_p=H^{-1}$, the amplitude of the correlation is $\langle\phi\phi\rangle\sim H^2$. On the super-horizon scale, the correlation does not decay and remains nearly constant. This value explains the primordial density fluctuations needed for the formation of large-scale structures in our present universe.
\begin{figure}[H]
\centering
\includegraphics[width=0.5\linewidth,clip]{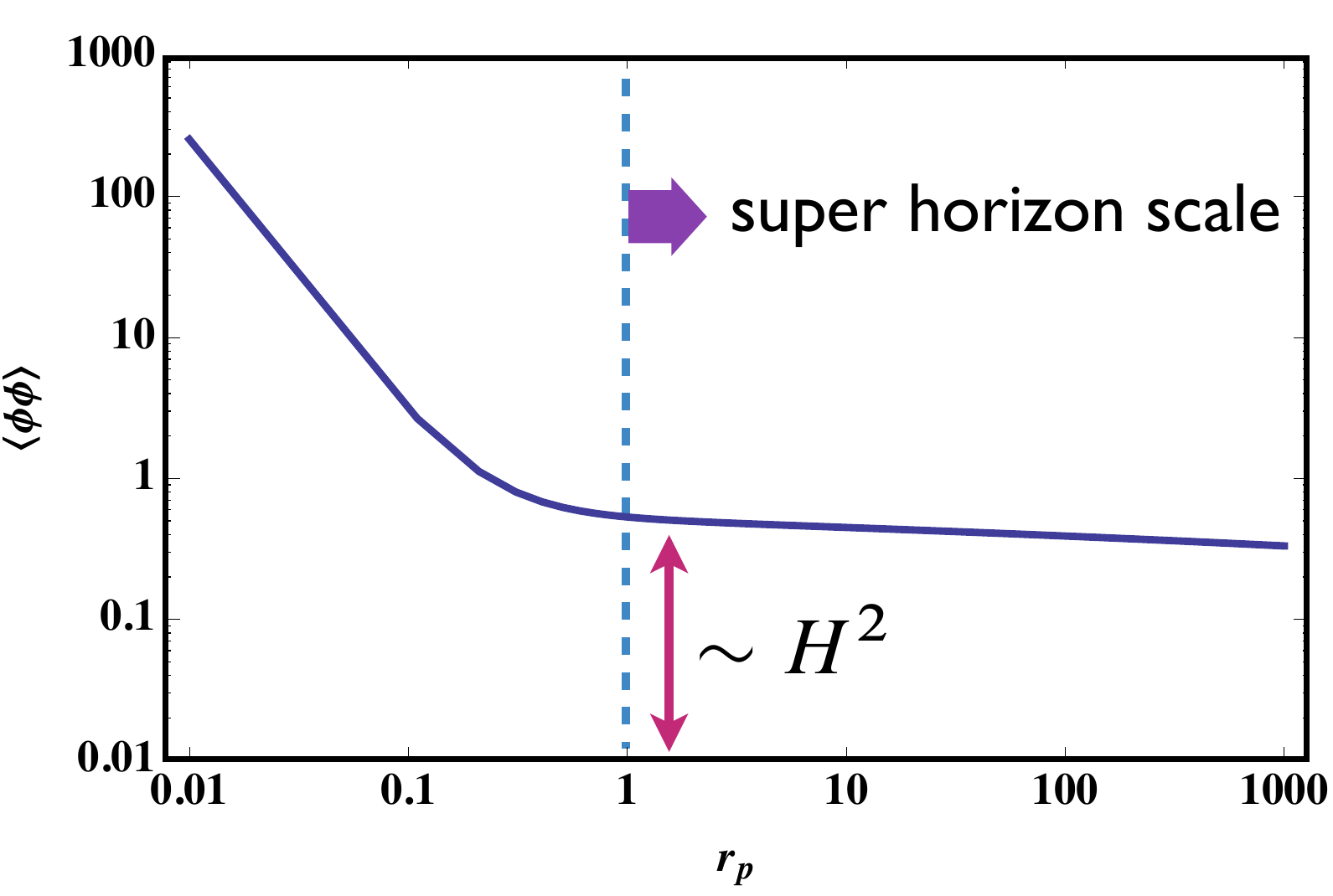}
\caption{Behavior of the equal time correlation function of the massless minimal scalar field ($H=1$). }
\end{figure}
\section{Behavior of the Negativity}
\vspace{-12pt}
\subsection{Causal Structure}

We explain our setup of the entanglement detection of the scalar field
using detectors (Figure 3). We consider an experiment that detects entanglement
of the scalar field using the detectors. The inflationary universe
begins at $t=t_0<0$ with physical size $H^{-1}$. At $t=0$, the size of
the universe is $k_0^{-1}$ where $k_0$ represents the infrared cut-off
in the Wightman function for the minimal scalar field. We fix the
value of the infrared cutoff $k_0$. For example, we take
$k_0/H=\exp(-20)$. Then we choose a specific value of the comoving
distance $r<k_0^{-1}$ between two detectors. This corresponds to
fixing the physical distance between the detectors at $t=0$. At $t=0$,
the switches of the detectors are ``on'' and the measurement of
entanglement is performed.
\begin{figure}[H]
\centering
\includegraphics[width=0.45\linewidth]{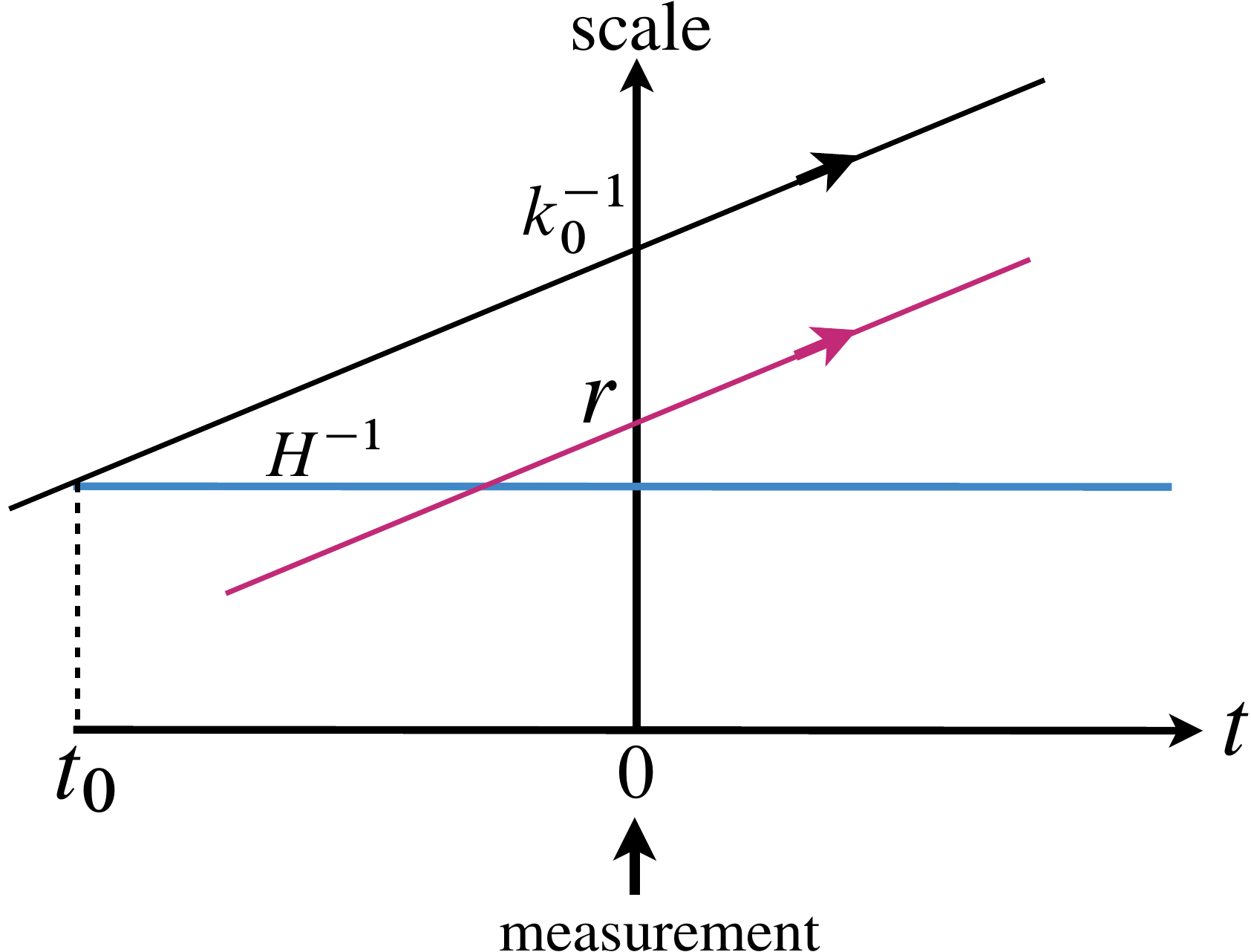}
\caption{ Setup of our thought experiment of detecting the entanglement in the inflationary universe.}
\end{figure}
In our analysis, the following Gaussian window function is adopted to represent ``on'' and ``off'' of~detectors:
\begin{equation}
\label{eq:window}
g(t)=g_0\exp\left(-\frac{t^2}{2\sigma^2}\right)
\end{equation}
The detector is ``on'' with duration $|t|\lesssim\sigma$ and ``off''
for the rest of time. This form of the window function approximates
the detector being ``on'' for $-\sigma\le t\le \sigma$. In the
Minkowski spacetime, the spacetime diagram with two detectors is
showin in Figure 4:
\begin{figure}[H]
\centering
\includegraphics[width=0.7\linewidth,clip]{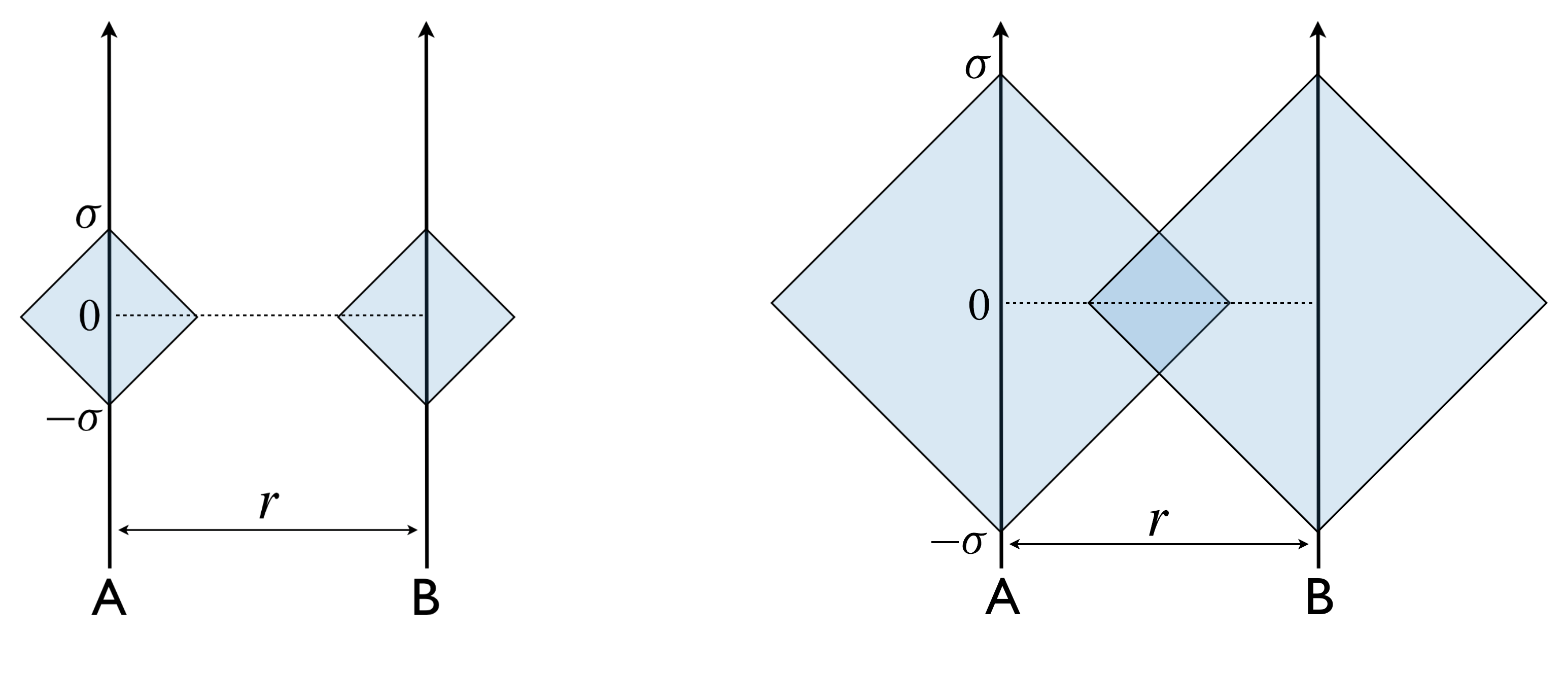}
\caption{Causal structure of two detectors system in the Minkowski spacetime. Shaded regions represent causal diamonds for each detector. Two detectors are causally disconnected for $r/\sigma>2$ (left panel). Two detectors are causally connected for $r/\sigma<2$ (right panel). }
\end{figure}
\noindent In the Minkowski spacetime, two detectors are spatially separated and causally disconnected for
\begin{equation}
\label{eq:spatial1}
2\sigma<r
\end{equation}
For two detectors with initial separable state satisfying this relation, we can detect the intrinsic entanglement of the scalar field because the entanglement between spatially disconnected regions cannot be produced via local process of measurement. On the other hand, for $r/\sigma<2$, causal interaction between two detectors is possible and it is unlikely to interpret the entanglement in this case as non-local correlations of the scalar field.

In de Sitter spacetime, the comoving size of the each causal diamond is
$$
r_{\text{null}}=\int_{-\sigma}^\sigma\frac{dt}{a}=\frac{2}{H}\sinh(H\sigma) $$
Thus the two comoving detectors are causally disconnected for
\begin{equation}
\label{eq:spatial2}
\frac{2}{H}\sinh(H\sigma)<r
\end{equation}
For the super-horizon scale $r>H^{-1}$, this condition reduces to
\begin{equation}
H\sigma\lesssim 0.48
\end{equation}
and for the sub-horizon scale $r<H^{-1}$, this condition reduces to
\begin{equation}
\begin{cases}
2\lesssim r/\sigma&\quad (H\sigma\lesssim 0.48) \\
\frac{1}{H\sigma}\lesssim r/\sigma&\quad (H\sigma\gtrsim 0.48)
\end{cases}
\end{equation}
By fixing the value of the Hubble constant $H$, the negativity is obtained as a function of $(r, \sigma,\Omega)$ in our~setup.

\subsection{$X$ and $E$}

To investigate the entanglement structure of the detectors system, we
must evaluate $X$ and $E$ defined by integrals \eqref{eq:XE}. Using
the Wightman function and the Gaussian window function
\eqref{eq:window}, they are 
\begin{align}
&E_{AB}=2g_0^2
\int_{-\infty}^\infty dx e^{-x^2/\sigma^2}
\int_{-\infty}^\infty dy e^{-y^2/\sigma^2-2i\Omega y}D^+(x,y,r) \notag\\
&E=2g_0^2
\int_{-\infty}^\infty dx e^{-x^2/\sigma^2}
\int_{-\infty}^\infty dy e^{-y^2/\sigma^2-2i\Omega y}D^+(x,y,0) \label{eq:XE2}\\
&X=-4g_0^2\int_{-\infty}^\infty dx e^{-x^2/\sigma^2+2i\Omega x}\int_0^\infty dy e^{-y^2/\sigma^2}
D^+(x,y,r) \notag
\end{align}
These quantities are necessary to obtain the negativity of the detectors system. For this purpose, using the contour integral on the complex plane and the method of residue, we rewrite the integral form of $E,X$ for numerical evaluation. The detail of the derivation is summarized in Appendix.

\subsubsection{Minkowski Vacuum and Thermal State}

For the Minkowski vacuum, $X,E$ can be obtained exactly: 
\begin{align}
&E=\frac{g_0^2}{4\pi}\left[e^{-\Omega^2\sigma^2}-\Omega\sigma\sqrt{\pi}\,
\mathrm{erfc}(\Omega \sigma)\right]\\
&X=-\frac{g_0^2\sigma}{4\sqrt{\pi}r}e^{-r^2/(4\sigma^2)-\Omega^2\sigma^2}
\left[\mathrm{erfi}\left(\frac{r}{2\sigma}\right)-i\right]
\end{align}
where {$$
\mathrm{erf}(x)=\frac{2}{\sqrt{\pi}}\int_0^x dt e^{-t^2},\quad\mathrm{erfc}(x)=\frac{2}{\sqrt{\pi}}\int_x^\infty dt e^{-t^2},\quad \mathrm{erfi}(x)=-i\,\mathrm{erf}(i\,x) $$}
For thermal state with temperature $T$, after evaluating contribution of poles in the Wightman function, we obtain the following integral formulas:
\begin{align}
&E=\frac{g_0^2}{2\pi}e^{-\Omega^2\sigma^2}\int_0^\infty dk ke^{-k^2}\frac{\cosh\left[\frac{\pi k}{h}(1-\frac{2}{\pi}h\Omega\sigma)\right]}
{\sinh\left[\frac{\pi k}{h}\right]}\\
&X=i\frac{g_0^2\sigma}{2\pi r}e^{-\Omega^2\sigma^2}\int_0^\infty dk e^{-k^2}\frac{\sinh\left[\frac{\pi k}{h}(1+i\frac{hr}{\sigma})\right]}{\sinh\left[
\frac{\pi k}{h}\right]}
\end{align}
where $h\equiv 2\pi T\sigma$.
\subsubsection{De Sitter}
For the massless minimal scalar field in de Sitter spacetime, after evaluating contributions of poles in the Wightman function by contour integration, we obtain $E=E_1+E_2, X=X_1+X_2$, where
\begin{align}
&E_1=\frac{g_0^2}{2\pi}e^{-\Omega^2\sigma^2}\int_0^\infty dkke^{-k^2}
\frac{\cosh\left[\frac{\pi k}{h}\left(1-\frac{2h\Omega\sigma}{\pi}\right)\right]
}{\sinh\left[\frac{\pi k}{h}\right]}\\
&X_1=\frac{ig_0^2}{4\pi^{3/2}}\int_{-\infty}^{\infty}dx
\frac{e^{-x^2+2i\Omega\sigma x}}{a\sqrt{a^2h^2+1}}\int_0^\infty dk e^{-k^2}
\frac{\sinh\left[\frac{\pi k}{h}\left(1+\frac{2i}{\pi}\ln b\right)\right]
}{\sinh\left[\frac{\pi k}{h}\right]}\\
&\qquad a=e^{hx}\frac{r}{2\sigma},\quad b=ah+\sqrt{a^2h^2+1},\quad h=H\sigma, \notag
\\
&E_2=-\frac{g_0^2h^2}{2\pi^2}\int_{-\infty}^\infty dxe^{-x^2}\int_{-\infty}^\infty dy e^{-y^2-2i\Omega\sigma y}\left\{\mathrm{Ei}\left[-2i\frac{k_0}{H}e^{-hx}\sinh(h(y-i\ep))\right]-1
\right\} \label{eq:E2}\\
&X_2=\frac{g_0^2h^2}{2\pi^2}\int_{-\infty}^\infty dx e^{-x^2+2i\Omega\sigma x}\int_0^\infty dy e^{-y^2} \notag \\
&\!\times\!\left\{\mathrm{Ei}\left[
-i\frac{k_0}{H}\left(-Hr+2e^{-hx}\sinh(h(y-i\ep))\right)\right]
\!+\!\mathrm{Ei}\left[-i\frac{k_0}{H}
\left(Hr\!+\!2e^{-hx}\sinh(h(y-i\ep))\right)\right]\!-\!2
\right\} \label{eq:X2}
\end{align}
$E_1$ and $X_1$ represent these quantities for the massless conformal scalar field. The negativity for conformal scalar is
\begin{equation}
\mathcal{N}=|X_1|-E_1
\end{equation}
The negativity for the minimal scalar is
\begin{equation}
\mathcal{N}=|X_1+X_2|-(E_1+E_2)
\end{equation}

\subsection{Asymptotic Analysis}

We can evaluate $x,y$ integral in \eqref{eq:XE} using the saddle point method~\cite{NambuY:PRD84:2011}. After rescaling the integration variables $x,y$,
\begin{align}
&E_{AB}=2g_0^2(\Omega\sigma^2)^2
\int_{-\infty}^\infty dx e^{-(\Omega\sigma)^2x^2}
\int_{-\infty}^\infty dy e^{-(\Omega\sigma)^2(y^2+2i y)}D^+(\Omega\sigma^2 x,\Omega \sigma^2y,r) \notag\\
&E=2g_0^2(\Omega\sigma^2)^2
\int_{-\infty}^\infty dx e^{-(\Omega\sigma)^2x^2}
\int_{-\infty}^\infty dy e^{-(\Omega\sigma)^2(y^2+2i y)}D^+(\Omega\sigma^2 x,\Omega\sigma^2 y,0) \\
&X=-4g_0^2(\Omega\sigma^2)^2\int_{-\infty}^\infty dx e^{-(\Omega\sigma)^2(x^2-2i\Omega x)}\int_0^\infty dy e^{-(\Omega\sigma)^2y^2}
D^+(\Omega\sigma^2 x,\Omega\sigma^2 y,r) \notag
\end{align}
We assume the following parameter ranges:
$$
\Omega\sigma\gg 1,\quad\frac{\pi}{h}\gg1,\quad\frac{r}{2\sigma}\gg 1 $$
These parameter ranges correspond to the analysis adopted in the paper~\cite{SteegGV:PRD79:2009}. The third condition means two detectors are causally disconnected. The second condition means that the separation of the poles in the complex $y$ plane is sufficiently large and the saddle points $x=0, y=-i$ provide main contribution to the $y$ integral of $E$. The third condition also means the separation between two poles $y=\pm r/2\sigma$ is sufficiently large and $y$ integral in $X$ can be evaluated by the saddle points $x=i, y=0$. The physical meaning of these conditions is that detectors are causally disconnected and the Hubble time scale is sufficiently larger than detector's duration time $\sigma$. Then, the asymptotic forms of integrals are
\begin{align}
&X\approx
-2\pi g_0^2\sigma^2e^{-(\Omega\sigma)^2}D^{+}(i\Omega\sigma^2, 0, r)
\\
&E\approx 2\pi g_0^2\sigma^2e^{-(\Omega\sigma)^2}D^{+}(0,
-i\Omega\sigma^2, 0)
\end{align}
Using these expressions, the negativity for the Minkowski vacuum is given by
\begin{equation}
\mathcal{N}\approx\frac{g_0^2}{8\pi}e^{-(\Omega\sigma)^2}
\left[\left(\frac{2\sigma}{r}\right)^2-\frac{1}{(\Omega\sigma)^2}\right]
\end{equation}
The negativity for the thermal state is
\begin{equation}
\mathcal{N}\approx 2\pi g_0^2\sigma^2e^{-\Omega^2\sigma^2}\times\frac{H^2}{16\pi^2}\left[
\frac{2}{Hr}\frac{\cosh(Hr/2)}{\sinh(Hr/2)}
-\frac{1}{\sin^2(h\Omega\sigma)}\right]
\end{equation}
For the conformal massless scalar field,
\begin{align*}
&X_1\approx -2\pi g_0^2\sigma^2e^{-\Omega^2\sigma^2}
\frac{H^2}{4\pi^2}\times\frac{e^{-2ih\Omega\sigma}}{(Hr)^2}\\
&E_{AB1}\approx 2\pi g_0^2\sigma^2e^{-\Omega^2\sigma^2}
\frac{H^2}{4\pi^2}\times\frac{1}{(Hr)^2+4\sin^2(h\Omega\sigma)}
\\
&E_1\approx 2\pi g_0^2\sigma^2e^{-\Omega^2\sigma^2}
\frac{H^2}{4\pi^2}\times\frac{1}{4\sin^2(h\Omega\sigma)}
\end{align*}
Thus, the negativity is
\begin{equation}
\mathcal{N}\approx 2\pi g_0^2\sigma^2e^{-\Omega^2\sigma^2}\times
\frac{H^2}{4\pi^2}\left[\frac{1}{(Hr)^2}-\frac{1}{4\sin^2(h\Omega\sigma)}
\right]
\end{equation}
We summarize the behaviors of negativities for these three types of scalar fields. In Figure~5, the negativity is positive in regions on the left side of the each $\mathcal{N}=0$ lines. In these regions, two detectors are entangled.
\begin{figure}[H]
\centering
\includegraphics[width=0.45\linewidth]{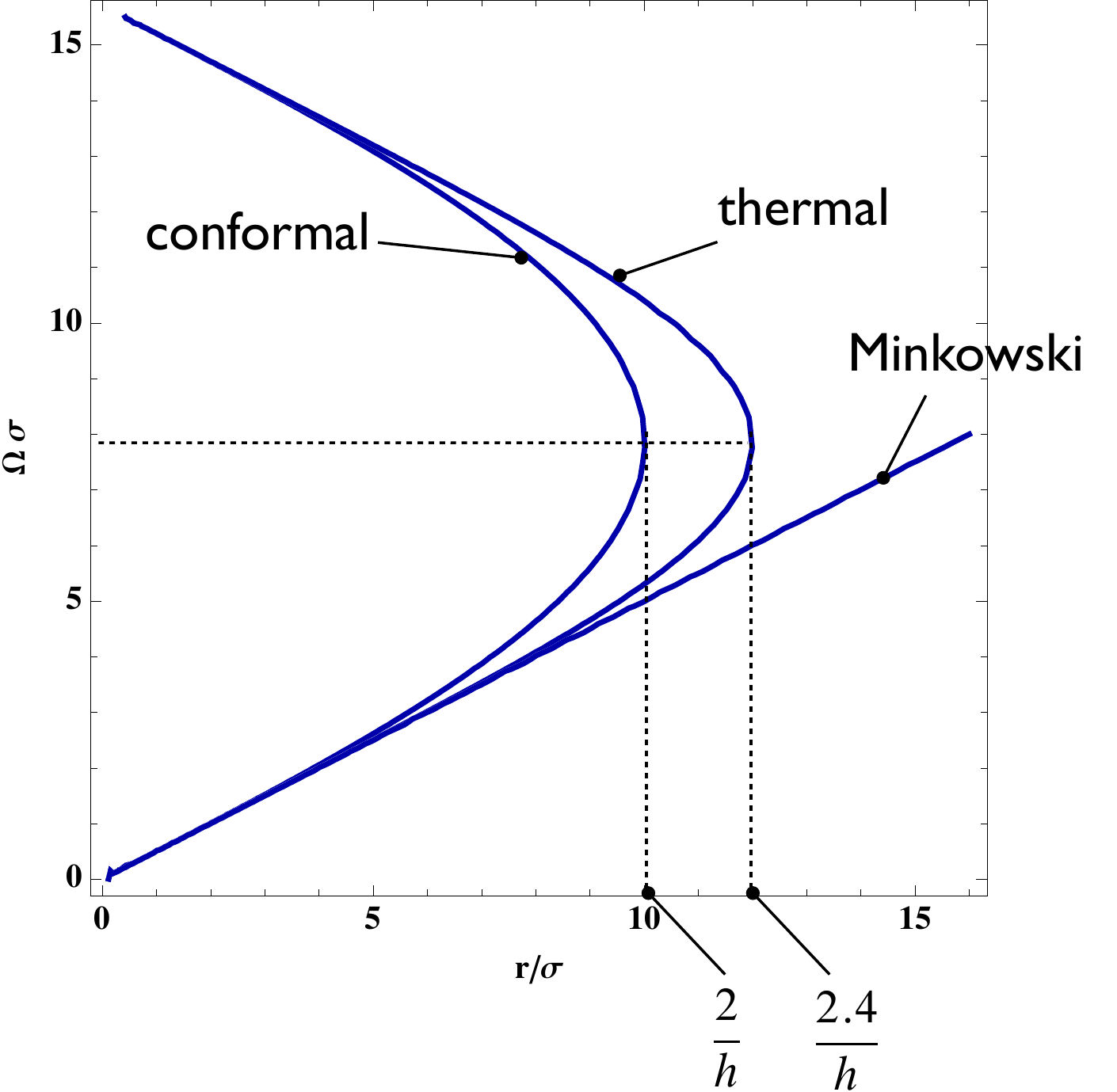}
\caption{$\mathcal{N}=0$ lines for $h=0.3$. Detectors are entangled in regions enclosed by $\mathcal{N}=0$ lines. The maximal spatial size of the entangled region is obtained for $\Omega\sigma=\pi/(2h)$.}
\end{figure}

For the Minkowski vacuum, it is always possible to find parameters $(r/\sigma, \Omega\sigma)$ with which the negativity is positive, which means that the Minkowski vacuum state is entangled. For the thermal state and the conformal vacuum state, the entangled regions in the parameter space become compact. Therefore, for sufficiently large separation of two detectors, we cannot detect positive negativity for any value of detector's parameter $\Omega,\sigma$. The maximal spatial size of the entangled region for the thermal state is $2.4H^{-1}$ and $2.0H^{-1}$ for the conformal scalar field. These behaviors reproduce the result of analysis done by G.~V.~Steeg and N.~C.~Menicucci~\cite{SteegGV:PRD79:2009}, which was obtained by asymptotic approximation with numerical estimation. The authors claim that the thermal nature of the quantum field due to cosmic expansions can be distinguished from the finite temperature effect in the Minkowski spacetime with the equivalent temperature.

For the minimal scalar field, we have the following additional terms in $X, E$ besides $X_1, E_1$:
\begin{align}
&X_2\approx 2\pi g_0^2\sigma^2e^{-(\Omega\sigma)^2}\left(\frac{H^2}{4\pi^2}\right)
\left[\ln(k_0 r)+\gamma-1\right]\\
&E_2\approx 2\pi g_0^2\sigma^2e^{-(\Omega\sigma)^2}\left(\frac{H^2}{4\pi^2}\right)
\left[-\ln\left(\frac{2k_0}{H}\sin(h\Omega\sigma)\right)-\gamma+1\right]
\notag
\end{align}
where we have assumed $k_0r\ll 1$ and $k_0/H\ll 1$. For $k_0/H\rightarrow 0$ limit, these quantities diverge. Using these asymptotic form of $X_{1,2}$ and $E_{1,2}$, the negativity for the detectors system interacting with the minimal scalar field is
\begin{equation}
\mathcal{N}\approx 2\pi g_0^2\sigma^2e^{-\Omega^2\sigma^2}
\frac{H^2}{4\pi^2}\left[
\ln\left(\frac{2\sin(h\Omega\sigma
)}{Hr}\right)+\frac{\cos(2h\Omega\sigma)}{(Hr)^2}
-\frac{1}{4\sin^2(h\Omega\sigma)}\right]+O\left(\frac{1}{\ln(k_0/H)}\right)
\end{equation}
For $k_0/H\rightarrow 0$, although the Wightman function has the infrared divergence and $X,E$ diverge, the negativity does not depend on the infrared cutoff $k_0$ and its value is finite. This infrared finiteness of the negativity was discussed in different context~\cite{MarcovitchS:PRA80:2009}. The maximal spatial size of the entangled region is obtained (see Figure~6)
\begin{equation}
Hr\approx 0.86 \quad \text{for}\quad h\Omega\sigma=\frac{\pi}{4}
\end{equation}
\begin{figure}[H]
\centering
\includegraphics[width=0.4\linewidth]{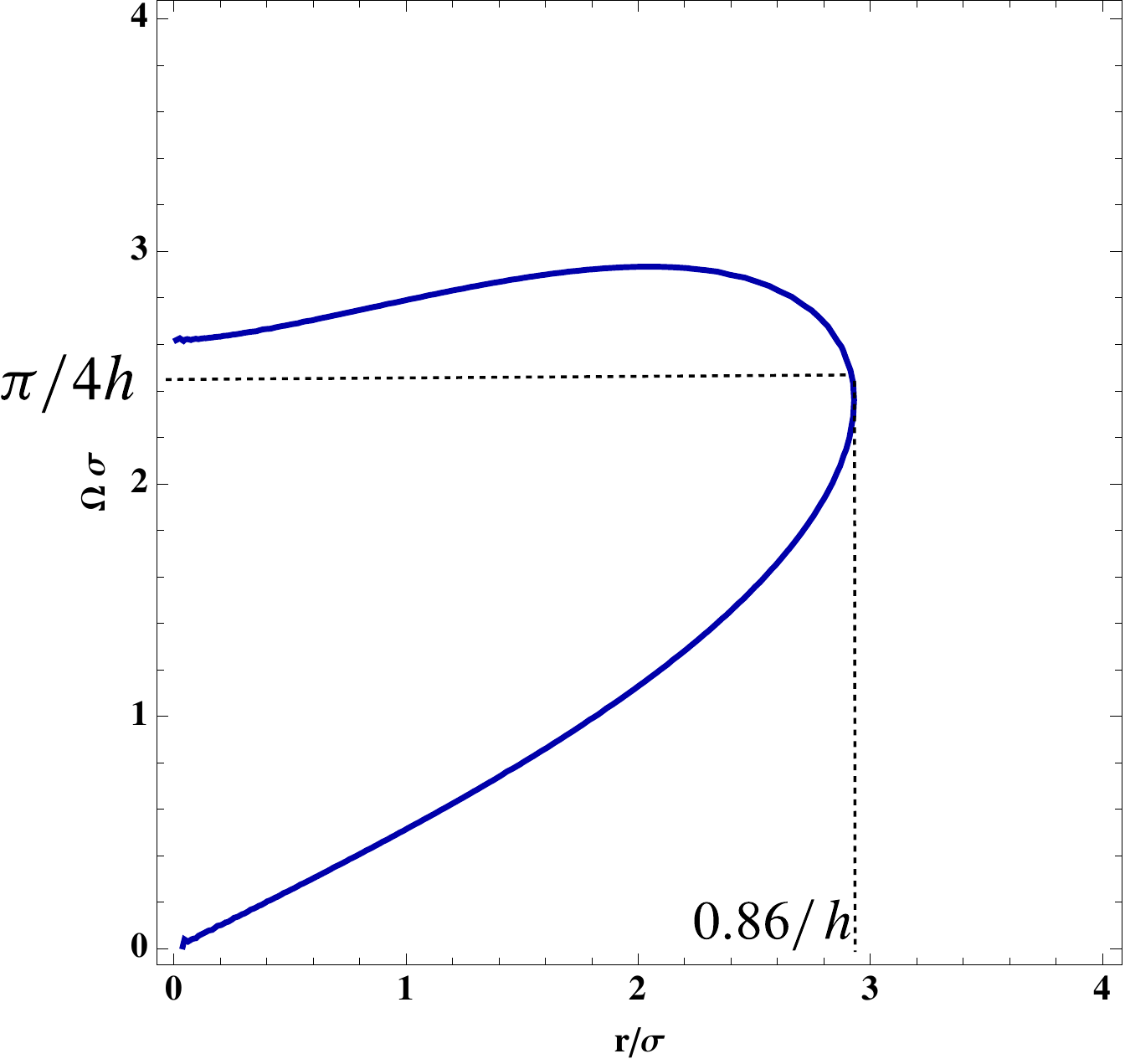}
\caption{$\mathcal{N}=0$ line for minimal scalar field for $h=0.3$. The scalar field is entangled in the region enclosed by the $\mathcal{N}=0$ line.}
\end{figure}
Thus, we do not detect the entanglement of the scalar field for the super-horizon scale $H^{-1}<r$. This result is contrasted with the result of the conformal scalar field, which has the maximal entangled size $2.0H^{-1}$ and the quantum correlation extends beyond the Hubble horizon scale.

\section{Numerical Evaluation of Negativity}

To confirm the result of the asymptotic analysis and investigate the behavior of the negativity for wider parameter ranges, we numerically calculated $X,E$ and the negativity.
\subsection{Thermal State}

For the thermal state, the numerical result is shown in Figure~7. The entangled region is ``compact'' in $(r/\sigma, \Omega\sigma)$ space and there exists the maximal distance $r_c$. Beyond that distance, two detectors do not catch the entanglement of the scalar field. This feature must be contrasted with the result for the Minkowski vacuum; in that case, the entangled region is not compact and the entanglement persists on all spatial~scales.
\begin{figure}[H]
\centering
\includegraphics[width=0.9\linewidth,clip]{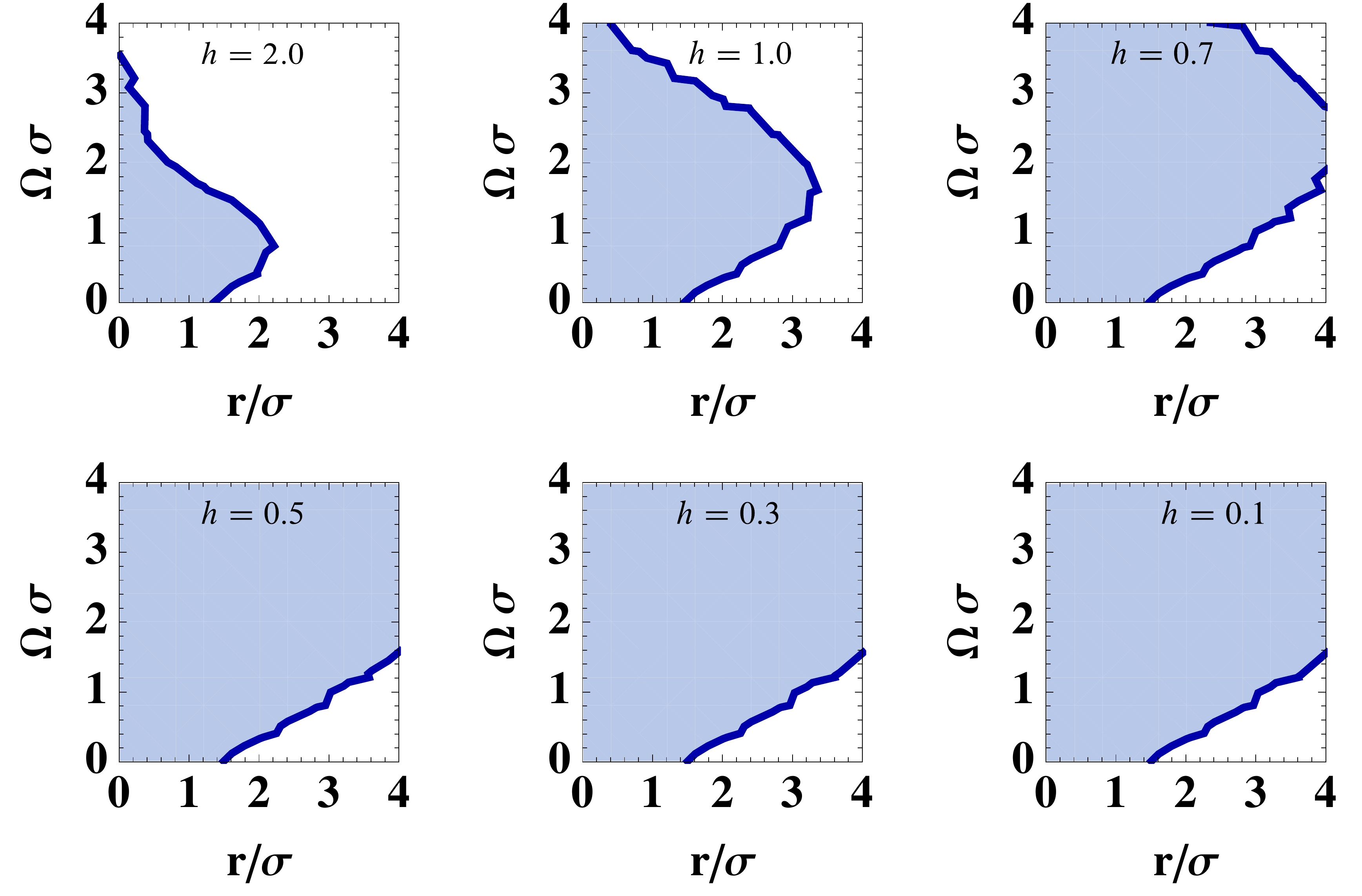}
\caption{The negativity of the massless scalar field with thermal state with temperature $T=H/2\pi$. Two detectors are entangled with parameters in the shaded regions. }
\end{figure}
The numerical results for $h<1$ are consistent with the asymptotic
analysis presented in Section~IV. For $h\gg 1$, $r_c H$ approaches
infinity as $\Omega/H\rightarrow 0$ ({Figure~8}).  
\begin{figure}[H]
\centering
\includegraphics[width=0.45\linewidth,clip]{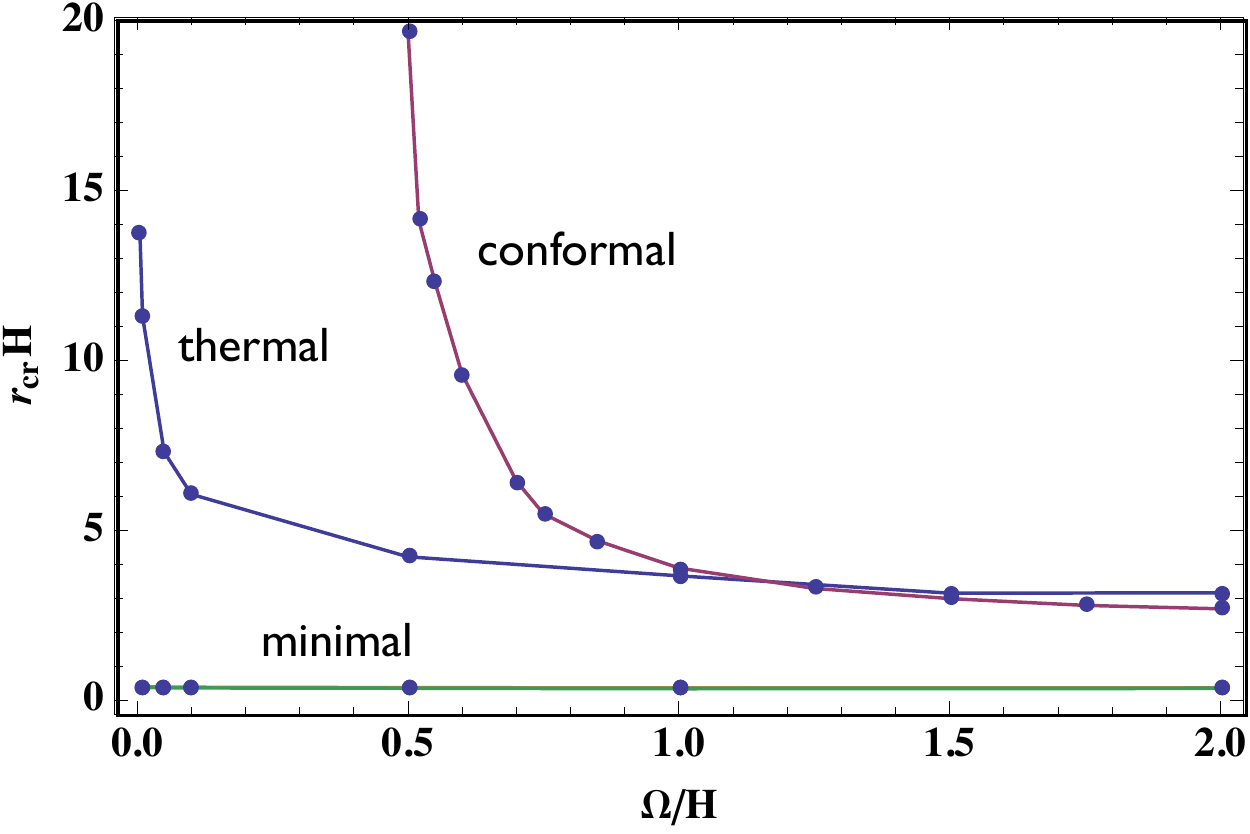}
\caption{The maximal spatial size $r_\text{c}$ of the entangled region as a function of $\Omega/H$. For a given value $\Omega/H$, the maximal entangled size $r_c$ is obtained in $(Hr, H\sigma)$ space.}
\end{figure}

\subsection{De Sitter}

This case corresponds to the scalar field in the inflationary universe. For the conformal scalar field, the result is shown in Figure~9.
\begin{figure}[H]
\centering
\includegraphics[width=0.9\linewidth,clip]{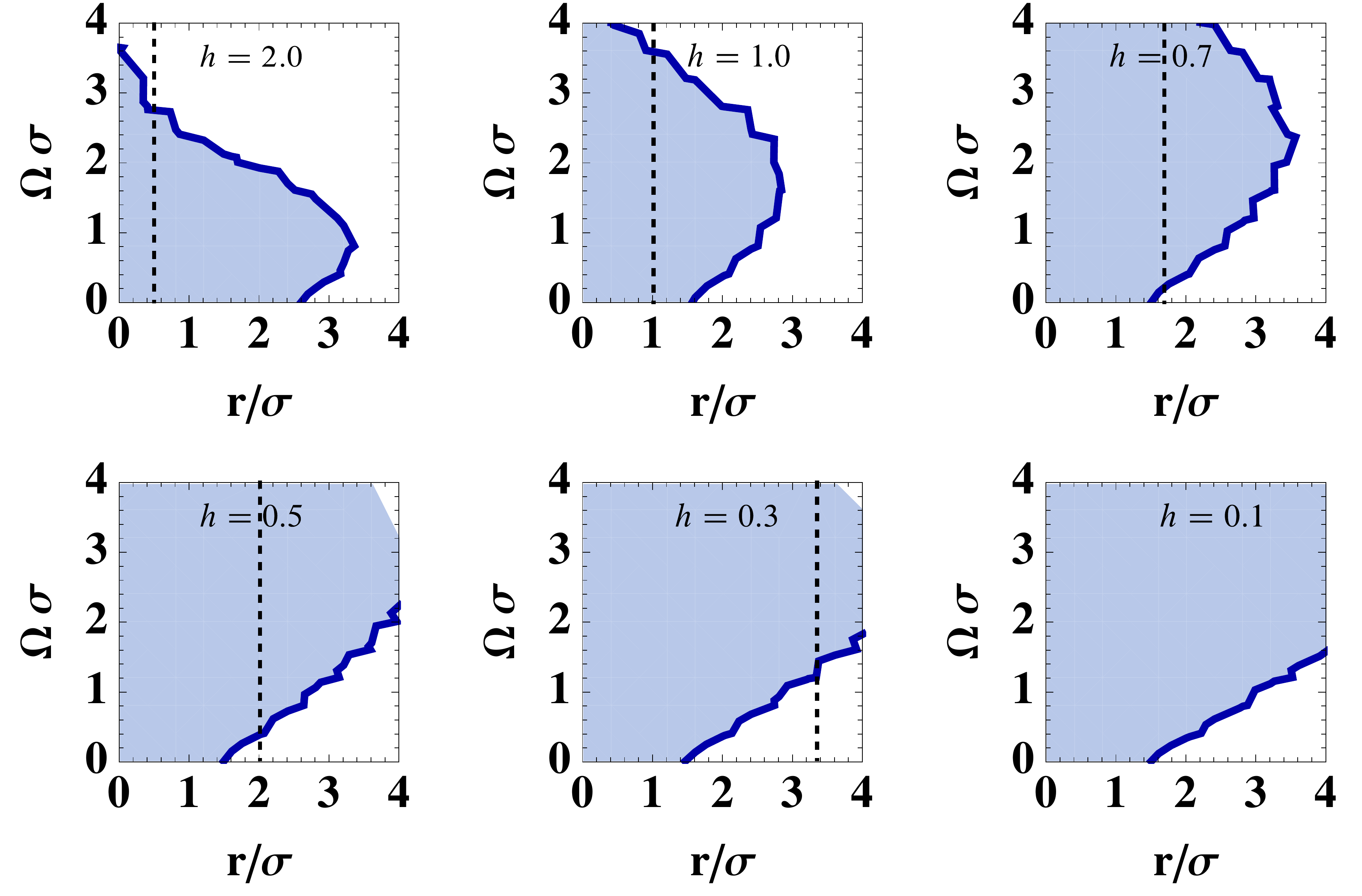}
\caption{The negativity of the massless conformal scalar field. Two detectors are entangled with parameters in the shaded region. The dotted lines represent the Hubble horizon scale $r=H^{-1}$.}
\end{figure}
The structure of entanglement for the massless conformal scalar field is as follows:
\begin{itemize}
\item For $h<1$, it is possible to detect entanglement of the scalar field for $r<2H^{-1}$. For larger separation $r>2H^{-1}$, the detectors are separable; however, this does not imply the scalar field is separable. Because the initial separable state evolves to the final separable state and we cannot say anything about the separability of the scalar field in this case.
\item For sufficiently large value of $h\gg 1$, two detectors can be entangled for any value of spatial separation (Figure~8). In this parameter regime, the maximal spatial size of the entangled region becomes infinite for $\Omega/H<1/2$. For $h\gg 1$, the causal diamonds of two detectors can have overlap even for $r>H^{-1}$ and two detectors are causally connected. Hence the entanglement between two detectors in this case does not mean non-local quantum correlation. The entanglement between detectors is established via causal contact between detectors.
\end{itemize}

For the minimal scalar field, the result is shown in {Figure~10}.
\begin{figure}[H]
\centering
\includegraphics[width=0.85\linewidth,clip]{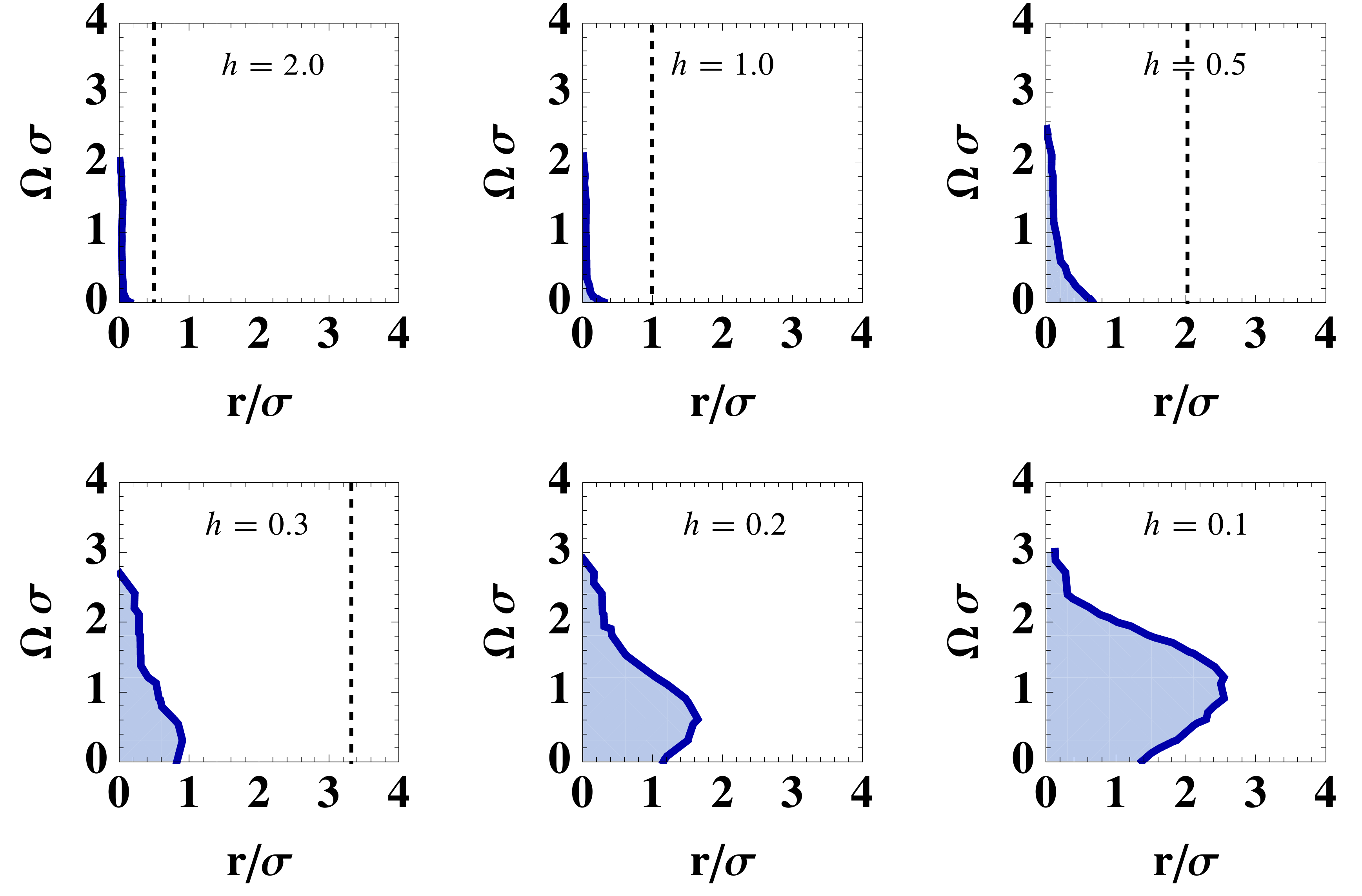}
\caption{The negativity of the massless minimal scalar field. Two detectors are entangled with parameters in the shaded region ($k_0/H=e^{-20}$). The dotted lines represent the Hubble horizon scale $r=H^{-1}$.}
\end{figure}
The structure of the entanglement for the massless minimal scalar field is as follows:
\begin{itemize}
\item The critical value $r_c$ is always about $r_c\sim 0.4H^{-1}$
and it is not possible to find out detector's parameters $\Omega,\sigma$ with which the two detectors are entangled beyond this scale. This result implies that the Bunch--Davies vacuum state for the massless minimal scalar field is not entangled for the large scale $r\gtrsim H^{-1}$ .
\item Our numerical evaluation of the negativity for the minimal scalar field strongly suggests that the minimal scalar field does not have the entanglement for the super-horizon scale. This is precisely our expected feature for classical behavior of the scalar field in the inflationary universe.
\end{itemize}

\section{Strength of Observable Correlation}

We can extract information on the correlation of the scalar field indirectly by analyzing the detectors' state. For this purpose, we consider the relation between the detectors' state and the observable (measurable) correlation between detectors in this section.

Using the Bloch representation~\cite{NielsenMA:CUP:2000}, the detectors' state \eqref{eq:bi-state} can be written as follows
\begin{equation}
\rho_{AB}=\frac{1}{4}\left( I\otimes I+\bs{a}\cdot\bs{\sigma}\otimes I+I\otimes\bs{b}\cdot\bs{\sigma}
+\sum_{\ell,m=1}^3 c_{\ell m}\sigma_\ell\otimes\sigma_m\right)
\end{equation}
where $\sigma_1, \sigma_2, \sigma_3$ are the Pauli matrices and the vectors $\bs{a}, \bs{b}$ and the matrix $c$ are defined by
\begin{align*}
&\bs{a}=\bs{b}=(0,0,-1+2E), \\
& c_{\ell m}=
\begin{pmatrix}
2(E_{AB}+\mathrm{Re}(X)) & -2\mathrm{Im}(X)& 0 \\
-2\mathrm{Im}(X) & 2(E_{AB}-\mathrm{Re}(X)) & 0 \\
0 & 0 & 1-4E
\end{pmatrix}
\end{align*}
We perform a local projective measurement of detectors' state and derive the information on the quantum correlation of the scalar field. Of course, it is not possible to perform measurement of the scalar field in the inflationary era directly. We consider the following measurement procedure as a gedanken experiment to explore the nature of quantum fluctuation of the scalar field. The measurement operators for each detector are
$$
M^{\pm}_A=\frac{I\pm\bs{n}_A\cdot\bs{\sigma}}{2},\quad M^{\pm}_B=\frac{I\pm\bs{n}_B\cdot\bs{\sigma}}{2},\quad|\bs{n}_A|=|\bs{n}_B|=1 $$
where $\pm$ denotes the output of the measurement. The vectors $\bs{n}_{A,B}$ represent the internal directions of the measurement for each detector. The joint probability $p_{jk}$ attaining measurement result $j$ for detector A and $k$ for detector B ($j,k=\pm 1$) is obtained as
\begin{align*}
p_{jk}&=\mathrm{tr}\left(M_A^j\otimes M_B^k\,\rho_{AB}\right)
=\frac{1}{4}\left[(1+(j)a_zn_z^A)(1+(k)b_zn_z^B)+(jk)
\sum_{\ell m}(c_{\ell m}-a_\ell b_m)n_\ell^An_m^B\right]\\
& c_{\ell m}-a_\ell b_m=
\begin{pmatrix}
2(E_{AB}+X_R) & -2X_I & 0 \\
-2X_I & 2(E_{AB}-X_R) & 0 \\
0 & 0 & 0
\end{pmatrix}
\end{align*}
The probability $p_j$ attaining a result $j$ for detector A and $p_k$
attaining a result $k$ for detector B are
$$
p_j=\sum_k p_{jk},\qquad p_k=\sum_j p_{jk}
$$

By the local projective measurement of the state, we obtain the following expectation values for qubit~variables:
$$
\langle\bs{n}_A\cdot\bs{\sigma}\rangle=\bs{a}\cdot\bs{n}_A,\quad
\langle\bs{n}_B\cdot\bs{\sigma}\rangle=\bs{b}\cdot\bs{n}_B,\quad
\langle\bs{n}_A\cdot\bs{\sigma}\otimes\bs{n}_B\cdot\bs{\sigma}\rangle
=\sum_{\ell,m}c_{\ell m}(n_A)_\ell(n_B)_m $$
For fluctuating parts of qubit variables $\Delta\sigma_{\bs{n}}=\bs{n}\cdot\bs{\sigma}-\langle\bs{n}\cdot\bs{\sigma}
\rangle$, the correlation function is
$$
\langle\Delta\sigma_{\bs{n}_A}\Delta\sigma_{\bs{n}_B}\rangle
=\sum_{\ell,m}(c_{\ell m}-a_\ell b_m)(n_A)_\ell(n_B)_m $$
For $\bs{n}_A=(\cos\theta_A,\sin\theta_A,0),
\bs{n}_B=(\cos\theta_B,\sin\theta_B, 0)$, the observable correlation of qubit variables is
\begin{equation}
\mathcal{C}\equiv
\langle\Delta\sigma_{\theta_A}\Delta\sigma_{\theta_B}\rangle=2(E_{AB}+|X|)
\cos\theta_{A}\cos\theta_{B}+2(E_{AB}-|X|)\sin\theta_{A}\sin\theta_{B}
\end{equation}
The amplitude of the correlation $\mathcal{C}$ depends on the behavior of $X,E_{AB}$ and the detectors' internal directions $\theta_A,\theta_B$. Using the result of the asymptotic estimation of $X,E_{AB}$,
\begin{align}
&X\approx-2\pi g_0^2\sigma^2e^{-(\Omega\sigma)^2}
\langle\phi(t+i\Omega\sigma^2,0)\phi(t+i\Omega\sigma^2,r)\rangle\\
&E_{AB}\approx 2\pi g_0^2\sigma^2e^{-(\Omega\sigma)^2}\langle\phi(t,0)
\phi(t+i\Omega\sigma^2,r)\rangle
\end{align}
The behaviors of these functions for the massless minimal scalar field are shown in Figure~11. The function $X$ is the equal time correlation function of the scalar field and reflects amplitude $H^2$ on the super-horizon~scale.
\begin{figure}[H]
\centering
\includegraphics[width=0.5\linewidth,clip]{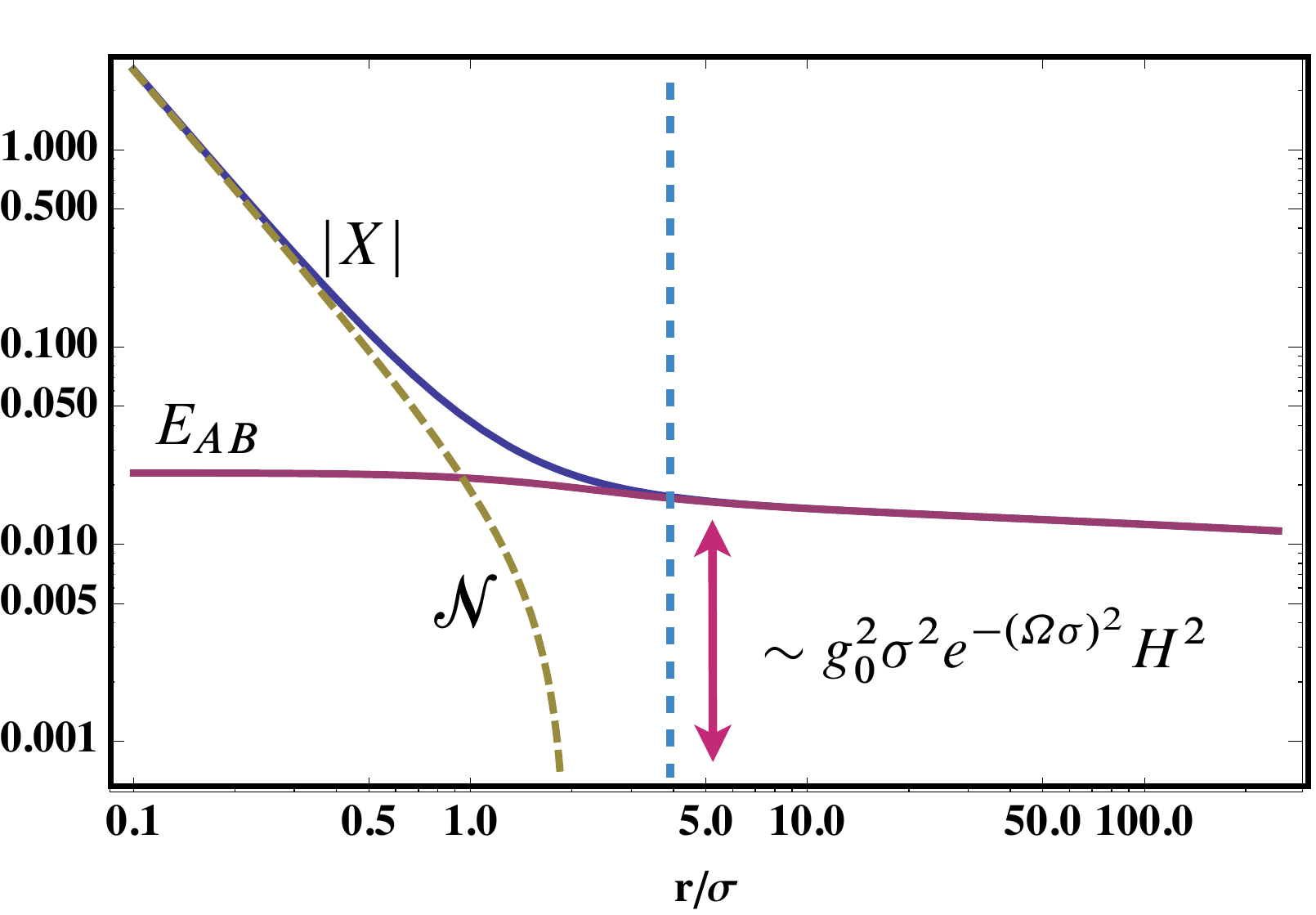}
\caption{The spatial dependence of $X, E_{AB}$ and the negativity $\mathcal{N}$ for the massless minimal scalar field. The vertical dotted line represents the Hubble horizon scale.}
\end{figure}
Let us accept that we do not know the internal angles of detectors $\theta_{A},\theta_{B}$ and consider the amplitude of the quantum correlation obtained by the measurement using these detectors. We first assume they are independent uniform random variables in $[0,2\pi]$. The distribution of the correlation $\mathcal{C}$ is shown in Figure~12 (left panel).

The values of the most probable correlations are
\begin{equation}
\mathcal{C}=E_{AB}+|X|, E_{AB}-|X|
\end{equation}
and they are realized with equal probability. Therefore, the expected amplitude of the correlation is
\begin{equation}
\label{eq:7}
\langle\mathcal{C}\rangle=E_{AB}
\end{equation}
This correlation explains the super-horizon scale correlation required for the structure formation.
\begin{figure}[H]
\centering
\includegraphics[width=0.45\linewidth,clip]{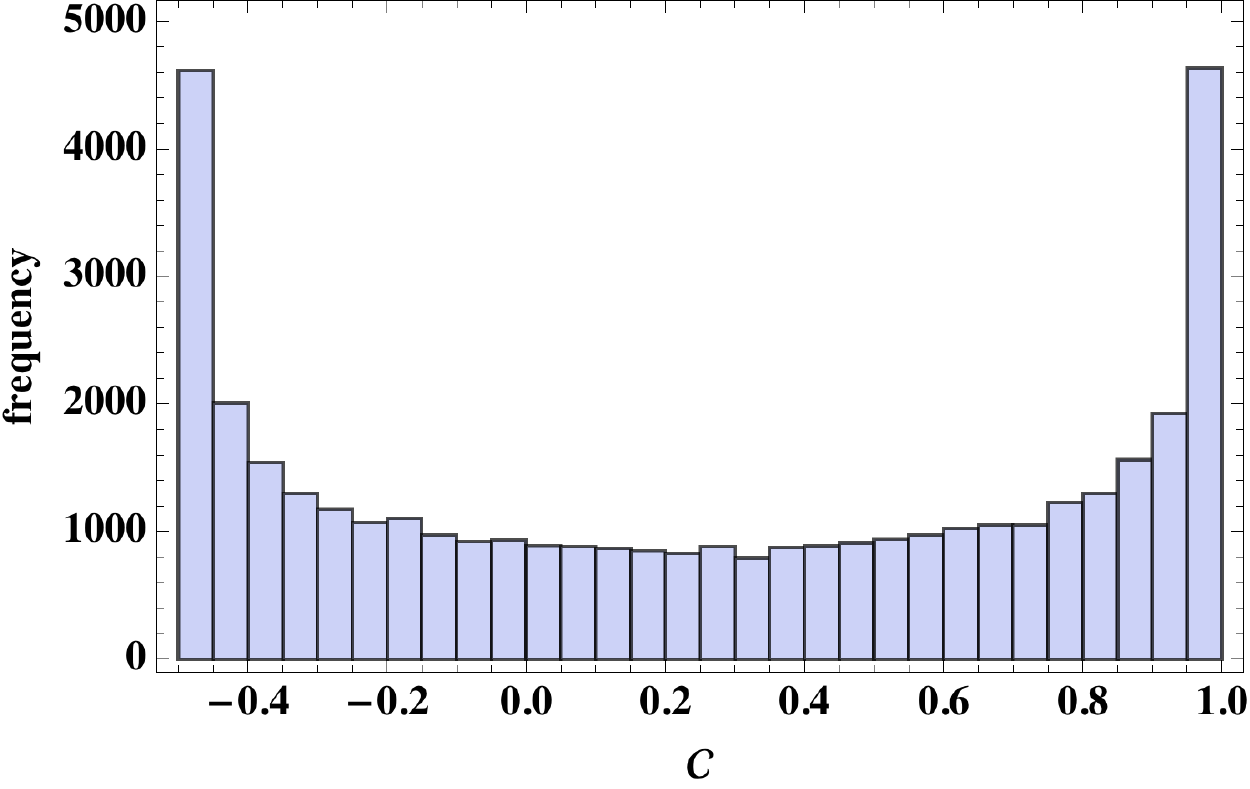}
\includegraphics[width=0.45\linewidth,clip]{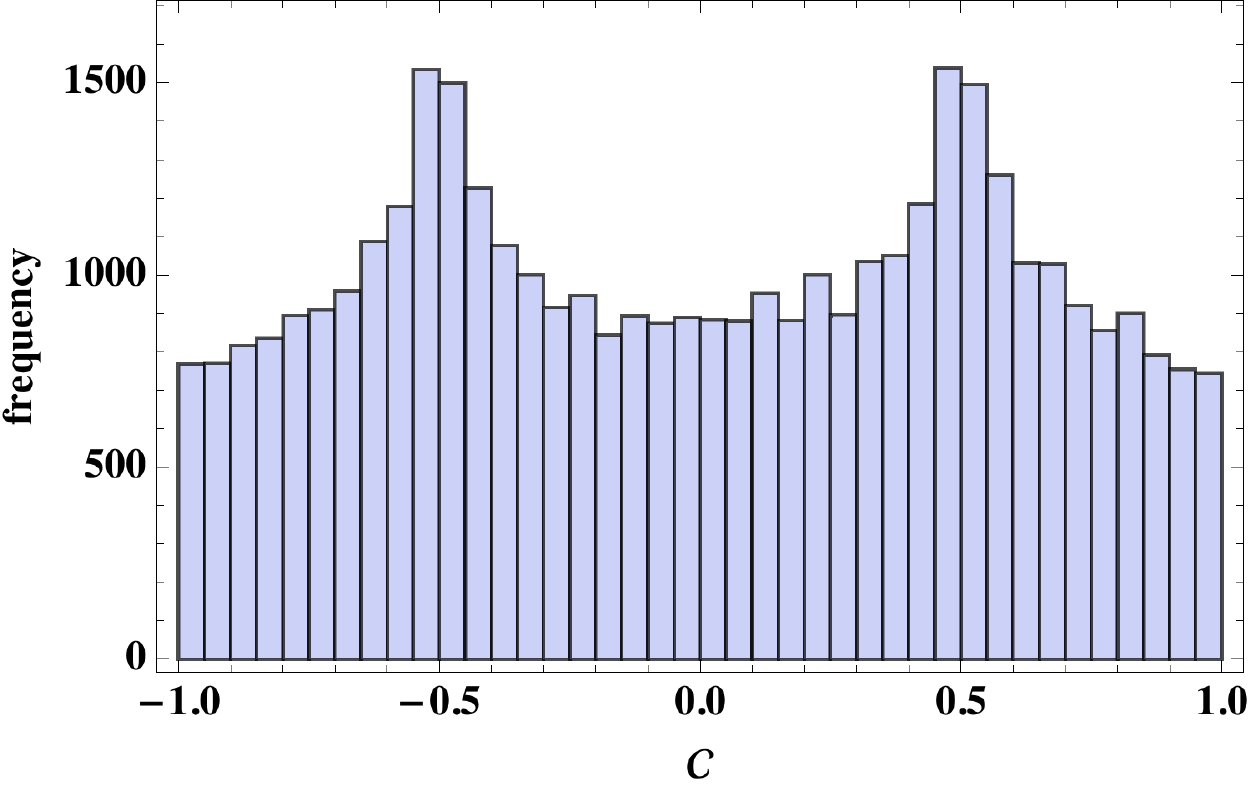}
\caption{Histogram for the correlation $\mathcal{C}$. Left panel: $\theta_A$ and $\theta_B$ are assumed to be independent random variables. The most probable correlation is $E_{AB}\pm
|X|$. Right panel: $\theta_A$ and $\theta_B$ are assumed to be random variables with $\theta_A=\theta_B$. The most probable correlation is $\pm(|X|-E_{AB})$.}
\end{figure}


Next, we consider other distribution of $\theta_A, \theta_B$. If we assume they are random variables with $\theta_A=\theta_B$, the most probable correlation is (see the right panel in Figure~12)
\begin{equation}
\mathcal{C}=\pm(|X|-E_{AB})
\end{equation}
and the least probable correlation is $\mathcal{C}=\pm(|X|+E_{AB})$. As is shown in Figure~13, the most probable correlation decays on large scale and cannot predict super-horizon scale correlation required for structure formation. Furthermore, the expected strength of correlation becomes $\langle\mathcal{C}\rangle=0$ in this case and we cannot acquire preferable correlation for primordial classical fluctuations. Measurable correlation strongly depends on the model of detector and the method of measurement. Thus we must specify the detail of detector models to predict observable correlation based on detector models.

\begin{figure}[H]
\centering
\includegraphics[width=0.6\linewidth,clip]{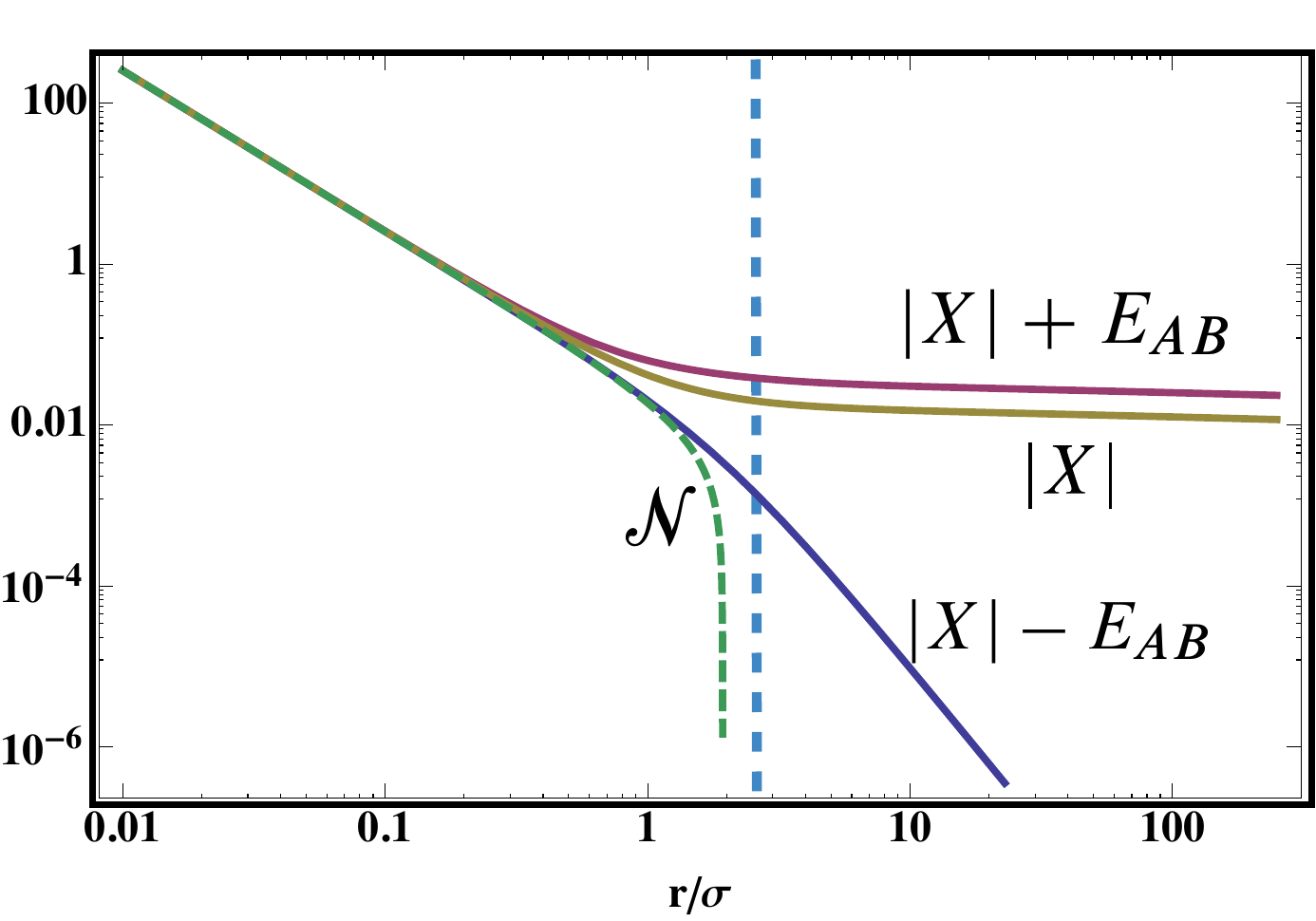}
\caption{The spatial dependence of $|X|\pm E_{AB}$ and $|X|$ for the massless minimal scalar field. The vertical dotted line represents the Hubble horizon scale.}
\end{figure}

\newpage
\section{Summary}

We investigated entanglement and quantum correlations of the quantum field in de Sitter spacetime using the detector model. Entanglement of the scalar field is swapped to two detectors interacting with the scalar field and we can measure the entanglement of the quantum field by this experimental setup. Entanglement structure of the scalar field depends on the type of the scalar field. For the massless conformal scalar field, we found that super-horizon scale entanglement is detected by suitably choosing detector's parameters. For the massless minimal scalar field, detectors cannot catch the entanglement of the scalar field on the super-horizon scale. This behavior is consistent with our previous analysis using the lattice model and the coarse-grained model of the scalar field~\cite{NambuY:PRD78:2008,NambuY:PRD80:2009}. In these analyses, by following the evolution of entanglement between two spatially separated regions, it was found the classicality of the quantum field appears as follows. Initially, when the size of the considered region is smaller than the Hubble horizon, the quantum field is in the entangled state. As the Universe expands, the quantum state becomes separable when the size of the region equals the size of the Hubble horizon. At this stage, the quantum correlation between neighboring regions is lost. Then, within about one Hubble time after the horizon crossing, noncommutativity of operators becomes negligible and the system can be treated as classical. In this article, we also discussed the outputs of detectors, which provide measurable or observable quantities, depend on the detail of internal structure of detectors. Thus, it is required to specify a reasonable model of detector and measurement process to predict nature of classical fluctuations originated from quantum fluctuations in the inflation.

As an application of our analysis presented in this paper, it is interesting to consider quantum effects in analogue curved spacetimes proposed using Bose--Einstein condensates or ion traps~\cite{BarceloC:0505065:2005}. In these experiential setups of analogue models, we can directly measure entanglement and classical and quantum correlations of quantum fluctuation using detectors in the laboratory. We expect that investigation in this direction will increase understanding of the quantum and classical nature generated during the inflation.

\section*{Acknowledgments}
\vspace{12pt}

The author would like to thank Yuji Ohsumi for valuable discussion on this subject. This article is based on works collaborated with him. This work was supported in part by the JSPS Grant-In-Aid for Scientific Research (C) (23540297).


\newpage
\section*{Appendix}

\section*{A. Contour Integration}
\vspace{12pt}

To calculate functions $X,E$ in the form of double integral numerically, we rewrite them by evaluating contributions of poles in the integrants. We present details of contour integrations for $X,E$.

\subsection*{A.1. Minkowski vacuum}
\vspace{12pt}

The Wightman function is
\begin{equation}
D^+=-\frac{1}{16\pi^2}\frac{1}{(y-i\ep)^2-r^2/4}
\end{equation}
The response function $E$ is
\begin{equation}
E=-\frac{g_0^2}{8\pi^{3/2}}
\int_{-\infty}^\infty dy\frac{e^{-y^2-2i\Omega\sigma y}}{(y-i\ep)^2}
\end{equation}
Using the identity
$$
e^{-y^2}=\frac{1}{\sqrt{\pi}}\int_{-\infty}^{\infty}dk e^{-k^2+2iky}
$$
$E$ can be written
$$
E=-\frac{g_0^2}{8\pi^2}
\int_{-\infty}^\infty dk e^{-(k+\Omega\sigma)^2}\int_{-\infty}^\infty dy
\frac{e^{2iky}}{(y-i\ep)^2}
$$
The $y$ integral
$$
I(k)=\int_{-\infty}^\infty dy
\frac{e^{2iky}}{(y-i\ep)^2}
$$
can be calculated using the contour $C_1C_2$ in the complex $y$ plane (Figure~A1).
\setcounter{figure}{0}
\renewcommand\thefigure{A\arabic{figure}}

\begin{figure}[H]
\centering
\includegraphics[width=0.4\linewidth]{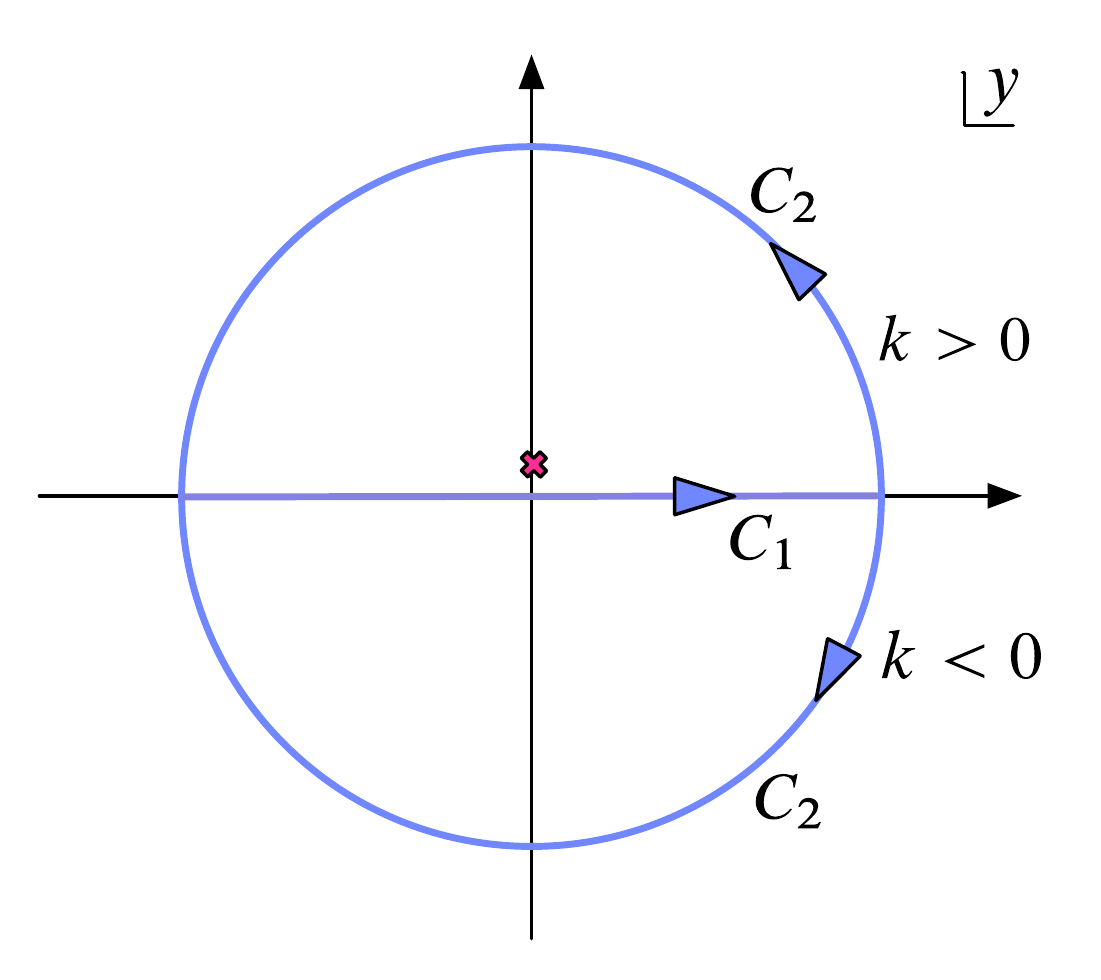}
\caption{The pole is $y=i\ep$. The radius of the contour $C_2$ is taken to be infinite.}
\end{figure}
\noindent The residue is
$$
\mathrm{Res}\left[\frac{e^{2iky}}{(y-i\ep)^2}\right]=2ik $$
For $k<0$, $I_{C_1}=0$ and for $k>0$, $I_{C_1}=2\pi i\times (2ik)=-4\pi k$. Thus,
\begin{align}
E&=\frac{g_0^2}{2\pi^2}\int_0^\infty dk e^{-(k+\Omega\sigma)^2}4\pi k \notag\\
&=\frac{g_0^2}{4\pi}\left[e^{-\Omega^2\sigma^2}-\sqrt{\pi}\,
\Omega\sigma\mathrm{erfc}(\Omega\sigma)\right]
\end{align}
$X$ is
\begin{align}
X&=\frac{g_0^2}{4\pi^{3/2}}e^{-\Omega^2\sigma^2}
\int_0^{\infty}\frac{e^{-y^2}}{(y-i\ep)^2-(r/(2\sigma))^2}\\
&=\frac{g_0^2}{4\pi^2}e^{-\Omega^2\sigma^2}
\int_{-\infty}^{\infty}dke^{-k^2}
\int_0^{\infty}dy\frac{e^{2iky}}{y^2-(r/(2\sigma))^2} \notag
\end{align}
We first evaluate $y$ integral with the contour $C_1C_2C_3$ (Figure~A2):
$$
I(k)\equiv\int_0^{\infty}dy\frac{e^{2iky}}{(y-i\ep)^2-(\frac{r}{2\sigma})^2}
=\frac{\sigma}{r}\int_0^{\infty}dy\left(\frac{1}{y-i\ep-\frac{r}{2\sigma}}
-\frac{1}{y-i\ep+\frac{r}{2\sigma}}\right) e^{2iky}
$$
\begin{figure}[H]
\centering
\includegraphics[width=0.4\linewidth]{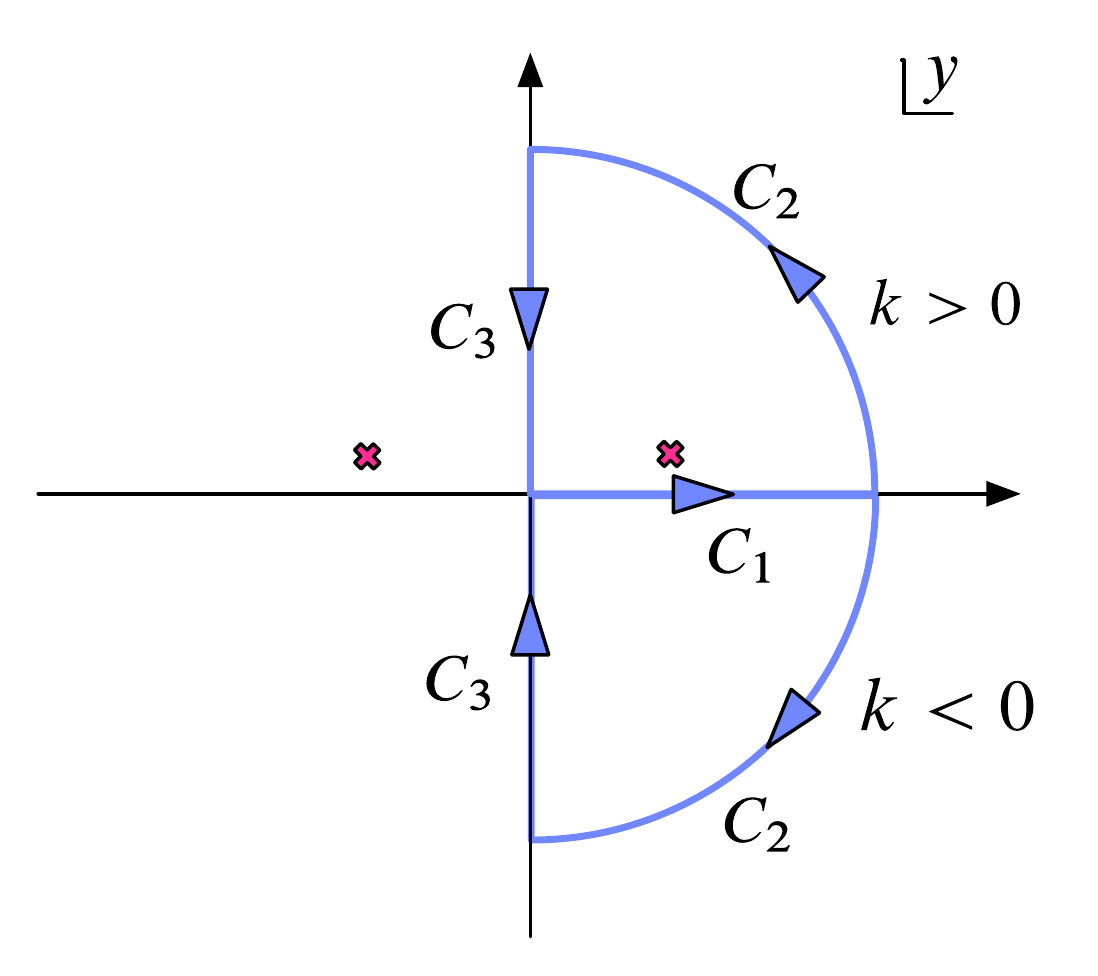}
\caption{Poles are $y=\pm r/(2\sigma)+i\ep$. The radius of the contour $C_2$ is taken to be infinite.}
\end{figure}
\noindent The residue for $y>0$ is
$$
\mathrm{Res}\left[\frac{e^{2iky}}{y-i\ep-\frac{r}{2\sigma}}\right]=e^{ikr/\sigma}
$$
By taking the radius of $C_2$ to infinity, $I_{C_2}=0$. For $k>0$,
$$
I_{C_1}+I_{C_3}=\frac{\sigma}{r}\times 2\pi i\times e^{ikr/\sigma}
$$
For $k<0$, $I_{C_1}+I_{C_3} =0$. On the other hand,
$$
I_{C_3}(k)=-i\mathrm{sign}(k)\int_0^{\infty}dy
\frac{e^{-2|k|z}}{y^2+(r/2\sigma)^2}
$$
This is an odd function of $k$ and does not contribute to $k$
integral. Thus, we obtain
\begin{align}
X&=\frac{ig_0^2\sigma}{2\pi r}e^{-\Omega^2\sigma^2}e^{2i\Omega t_0}
\int_{0}^{\infty}dk e^{-k^2+irk/\sigma} \notag \\
&=-\frac{g_0^2\sigma}{4\sqrt{\pi}r}e^{-\Omega^2\sigma^2}e^{-r^2/(4\sigma^2)}
e^{2i\Omega t_0}
\left[\mathrm{erfi}(r/(2\sigma))-i\right]
\end{align}

\subsection*{A.2. The Thermal State}
\vspace{12pt}

For the thermal state, the Wightman function is
\begin{equation}
D_T^+(x,y,r)=\frac{1}{8\pi r\beta}\left[\coth\frac{\pi(r-2y+i\ep)}{\beta}
+\coth\frac{\pi(r+2y-i\ep)}{\beta}\right],\quad \beta=\frac{1}{T}
\end{equation}
and
$$
D_T^+|_{r=0}=-\frac{T^2}{4}\frac{1}{\sinh^2(2\pi T (y-i\ep))}=-\frac{H^2}{16\pi}\frac{1}{\sinh^2(H(y-i\ep))},
\quad H\equiv 2\pi T $$
The response function is
\begin{align}
E&=-\left(\frac{g_0^2h^2}{8\pi^{3/2}}\right)\int_{-\infty}^{\infty}dy
\frac{e^{-y^2-2i\Omega\sigma y}}{\sinh^2(h(y-i\ep))},\qquad h\equiv H\sigma\\
&=-\frac{g_0^2}{8\pi^2}\int_{-\infty}^{\infty}dk e^{-(k+\Omega\sigma)^2}\int_{-\infty}^{\infty}dy\sum_{n=-\infty}^{\infty}
\frac{e^{2iky}}{\left(y-i\ep-i\frac{\pi n}{h}\right)^2} \notag
\end{align}
The integral
$$
I_1(k)= \int_{-\infty}^{\infty}dy\sum_{n=-\infty}^{\infty}
\frac{e^{2iky}}{(y-i\ep-i\pi n/h)^2}
$$
can be calculated using the contour $C_1C_2$ in the complex $y$ plane (Figure~A3).
\begin{figure}[H]
\centering
\includegraphics[width=0.4\linewidth,clip]{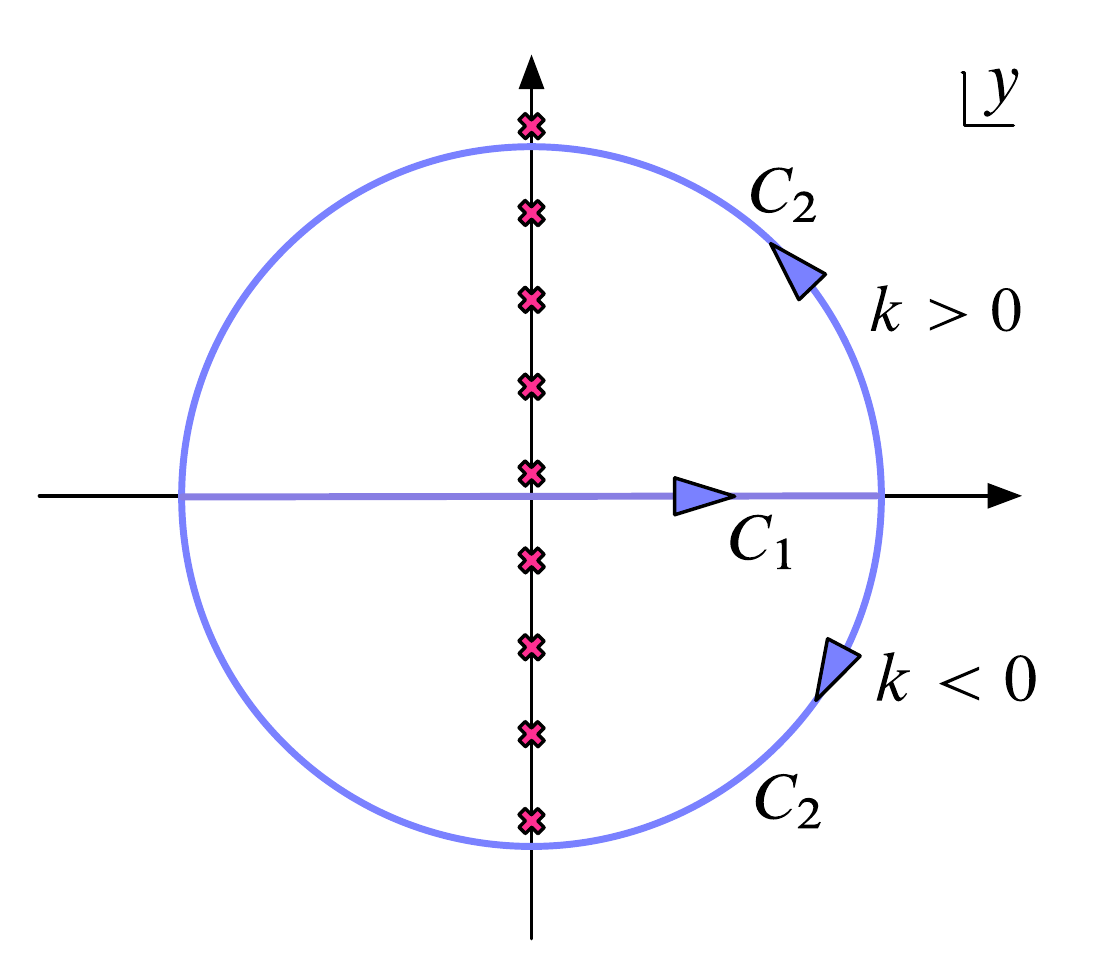}
\caption{The integration contour for $I_1(k)$. The radius of the contour $C_2$ is taken to be~infinite.}
\end{figure}
\noindent The residue for each pole is
$$
\mathrm{Res}\left[\frac{e^{2iky}}{(y-i\pi n/h)^2}\right]=2ik e^{-2\pi kn/h}
$$
and collecting all contribution from poles in the closed contour, the integral becomes
\begin{align*}
&I_1(k>0)=2\pi i\times\sum_{n=0}^{\infty}2ike^{-2\pi kn/h}=-\frac{4\pi k}{1-e^{-2\pi k/h}} \\
&I_1(k<0)=-2\pi i\times\sum_{n=-\infty}^{-1}2ike^{-2\pi kn/h}=-\frac{4\pi k}{1-e^{-2\pi k/h}} \notag
\end{align*}
Thus, the response function is obtained as
\begin{align}
E &=\frac{g_0^2}{2\pi}\int_{-\infty}^{\infty}dk e^{-(k+\Omega\sigma)^2}
\frac{k}{1-e^{-2\pi k/h}}\notag \\
&=\frac{g_0^2}{2\pi}e^{-\Omega^2\sigma^2}\int_{0}^{\infty}dk ke^{-k^2}
\frac{\cosh\left[\frac{\pi k}{h}\left(1-\frac{2\Omega\sigma h}{\pi}\right)\right]}{\sinh\left[\frac{\pi k}{h}\right]},\quad h=2\pi T\sigma \label{eq:app-thermal2}
\end{align}
$X$ is
\begin{equation}
X=-4g_0^2\sigma^2e^{-\Omega^2\sigma^2}\int_{-\infty}^{\infty}
dk e^{-k^2}\int_0^{\infty}dy e^{2iky}D_T^{+}(\sigma y)
\end{equation}
We first evaluate $y$ integral using contour integral:
\begin{align*}
I_2(k)&=\int_0^\infty dy e^{2iky}D_T^{+}(\sigma y) \notag \\
&=\frac{1}{16\pi^2 r\sigma}\int_0^\infty dy e^{2iky}\sum_{n=-\infty}^{\infty}
\left[\frac{-1}{y-i\ep-\left(\frac{r}{2\sigma}+\frac{in}{2T\sigma}\right)}+
\frac{1}{y-i\ep-\left(-\frac{r}{2\sigma}+\frac{in}{2T\sigma}\right)}\right]
\end{align*}

\noindent We use the contour $C_1C_2C_3$. The residue for poles in $y>0$ is
$$
\mathrm{Res}\left[\frac{e^{2iky}}{y-\left(\frac{r}{2}
+\frac{in}{2T\sigma}\right)}\right]= e^{ikr/\sigma}e^{-kn/(T\sigma)}
$$
For $k>0$,
$$
I_2(C_1)+I_2(C_3)=\frac{1}{16\pi^2r\sigma}\times 2\pi i\times\sum_{n=0}^\infty e^{ikr/\sigma}e^{-kn/(T\sigma)}=-\frac{i}{8\pi r}\frac{e^{ikr/\sigma}}{1-e^{-k/(T\sigma)}}
$$
For $k<0$,
$$
I_2(C_1)+I_2(C_3)=\frac{1}{16\pi^2r\sigma}\times (-2\pi i)
\times\sum_{n=-\infty}^{-1}
e^{ikr/\sigma}e^{-kn/(T\sigma)}=-\frac{i}{8\pi r}\frac{e^{ikr/\sigma}}{1-e^{-k/(T\sigma)}}
$$
As $I_2(C_3)_{k>0}=-I_2(C_3)_{k<0}$, the integral $I_2(C_3)$ does not contribute to $X$ after $k$ integration. Therefore,
\begin{align}
X &=i\frac{g_0^2\sigma}{2\pi r}e^{-\Omega^2\sigma^2}
\int_{-\infty}^{\infty}dk\frac{e^{-k^2+ikr/\sigma}}{1-e^{-k/(T\sigma)}}
\notag \\
&=i\frac{g_0^2\sigma}{2\pi r}e^{-\Omega^2\sigma^2}
\int_0^{\infty}dk e^{-k^2}\frac{\sinh\left[\frac{\pi k}{h}\left(1+\frac{irh}{\pi\sigma}\right)\right]}{\sinh\left[\frac{\pi k}{h}\right]},\quad h=2\pi T\sigma
\end{align}
\subsection*{A.3. The Conformal Scalar Field}
\vspace{12pt}

The integrals are
\begin{align}
&E_1=2\sqrt{\pi}\,g_0^2\sigma
\int_{-\infty}^{\infty}dy e^{-(y/\sigma)^2-2i\Omega y}D_C^{+}(x,y)|_{r=0} \notag \\
&X_1=4g_0^2\int_{-\infty}^{\infty}dx e^{-(x/\sigma)^2+2i\Omega x}\int_0^{\infty}dy e^{-(y/\sigma)^2}D_C^{+}(x,y) \\
& D_C^{+}(x,y)=-\frac{H^2}{4\pi^2}\left[4\sinh^2(H(y-i\ep))
-e^{2Hx}H^2r^2\right]^{-1}\notag
\end{align}
The response function is
\begin{align}
E_1&=-\frac{g_0^2h^2}{8\pi^{3/2}}\int_{-\infty}^{\infty}dy
\frac{e^{-y^2-2i\Omega\sigma y}}{\sinh^2(h(y-i\ep))},\quad h=H\sigma \notag
\\
&=-\frac{g_0^2}{8\pi^2}\int_{-\infty}^{\infty}dk e^{-(k+\Omega\sigma)^2}
\int_{-\infty}^{\infty}dy\sum_{n=-\infty}^{\infty}
\frac{e^{2iky}}{(y-i\ep-i\pi n/h)^2}
\end{align}
This is the same as the thermal state with $h=H\sigma$ and the result of the integration is given by \eqref{eq:app-thermal2} with $h=H\sigma$.

The $X_1$ is
\begin{equation}
X_1=\frac{g_0^2}{4\pi^{5/2}}\int_{-\infty}^{\infty}dx e^{-x^2+2i\Omega\sigma x}\int_{-\infty}^{\infty}dk e^{-k^2}\int_0^{\infty}\frac{e^{2iky}}{(\sinh h(y-i\ep)/h)^2-e^{2hx}(r/(2\sigma))^2}
\end{equation}
The integral
\begin{align*}
&I_2(k)=\int_0^{\infty}\frac{e^{2iky}}{(\sinh h(y-i\ep)/h)^2-a^2}\notag \\
&=\frac{1}{2a\sqrt{a^2h^2+1}}\int_0^{\infty}dy e^{2iky}\sum_{n=-\infty}^{\infty}
\left[\frac{1}{y-i\ep-(\ln b/h+in\pi/h)}-\frac{1}{y-i\ep-(-\ln b/h+in\pi/h)}\right]\\
&\quad a=e^{hx}r/(2\sigma),\quad b=a+\sqrt{a^2+1} \notag
\end{align*}
can be calculated using the contour $C_1C_2C_3$ in the complex $y$
plane (Figure~A4). The residue of poles for $y>0$ is
$$
\mathrm{Res}\left[\frac{e^{2iky}}{y-(\ln b/h+in\pi/h)}\right]=e^{(2ik/h)\ln b}e^{-(2\pi k/h)n}
$$
For $k>0$, the contribution of the contours $C_1,C_3$ to the integral is
\begin{align*}
I_2(C_1)+I_2(C_3)&=2\pi i\times\sum_{n=0}^{\infty}e^{(2ik/h)\ln b}e^{-(2\pi k/h)n}\times\frac{1}{2a\sqrt{a^2h^2+1}}\notag \\
&=\frac{2\pi ie^{(2ik/h)\ln b}}{1-e^{-2\pi k/h}}\frac{1}{2a\sqrt{a^2h^2+1}}
\end{align*}
For $k<0$,
\begin{align*}
I_2(C_1)+I_2(C_3)&=-2\pi i\times\sum_{n=-1}^{-\infty}e^{(2ik/h)\ln b}e^{-(2\pi k/h)n}\times\frac{1}{2a\sqrt{a^2h^2+1}}\notag \\
&=\frac{2\pi ie^{(2ik/h)\ln b}}{1-e^{-2\pi k/h}}\frac{1}{2a\sqrt{a^2h^2+1}}
\end{align*}
The contribution of the contour $C_3$ is
$$
I_2(C_3)
=i\mathrm{sign}(k)\int_0^{\infty}dy\frac{e^{-2|k|y}}{(\sinh hy/h)^2+a^2}
$$
This is an odd function with respect to $k$ and does not contribute to $X_1$ after the $k$ integration. We finally obtain
\begin{align}
X_1
&=\frac{ig_0^2}{4\pi^{3/2}}\int_{-\infty}^{\infty}dx\frac{e^{-x^2+2i\Omega\sigma x}}{a\sqrt{a^2h^2+1}}\int_{-\infty}^{\infty}dk\frac{e^{-k^2+(2i\ln b/h)k}}{1-e^{-2\pi k/h}} \notag \\
&=\frac{ig_0^2}{4\pi^{3/2}}\int_{-\infty}^{\infty}dx\frac{e^{-x^2+2i\Omega\sigma x}}{a\sqrt{a^2h^2+1}}\int_0^{\infty}dk e^{-k^2}
\frac{\sinh\left[\frac{\pi k}{h}\left(1+\frac{2i}{\pi}\ln b\right)\right]}{
\sinh\left[\frac{\pi k}{h}\right]}
\end{align}
If the limit $h\rightarrow 0$ is taken in (75) and (77), we obtain the expression for the Minkowski vacuum.

For the minimal scalar field, we also have to calculate the integrals $X_2, E_2$ given by \eqref{eq:E2} and \eqref{eq:X2}. They are evaluated numerically in that form.

\begin{figure}[H]
\centering
\includegraphics[width=0.4\linewidth,clip]{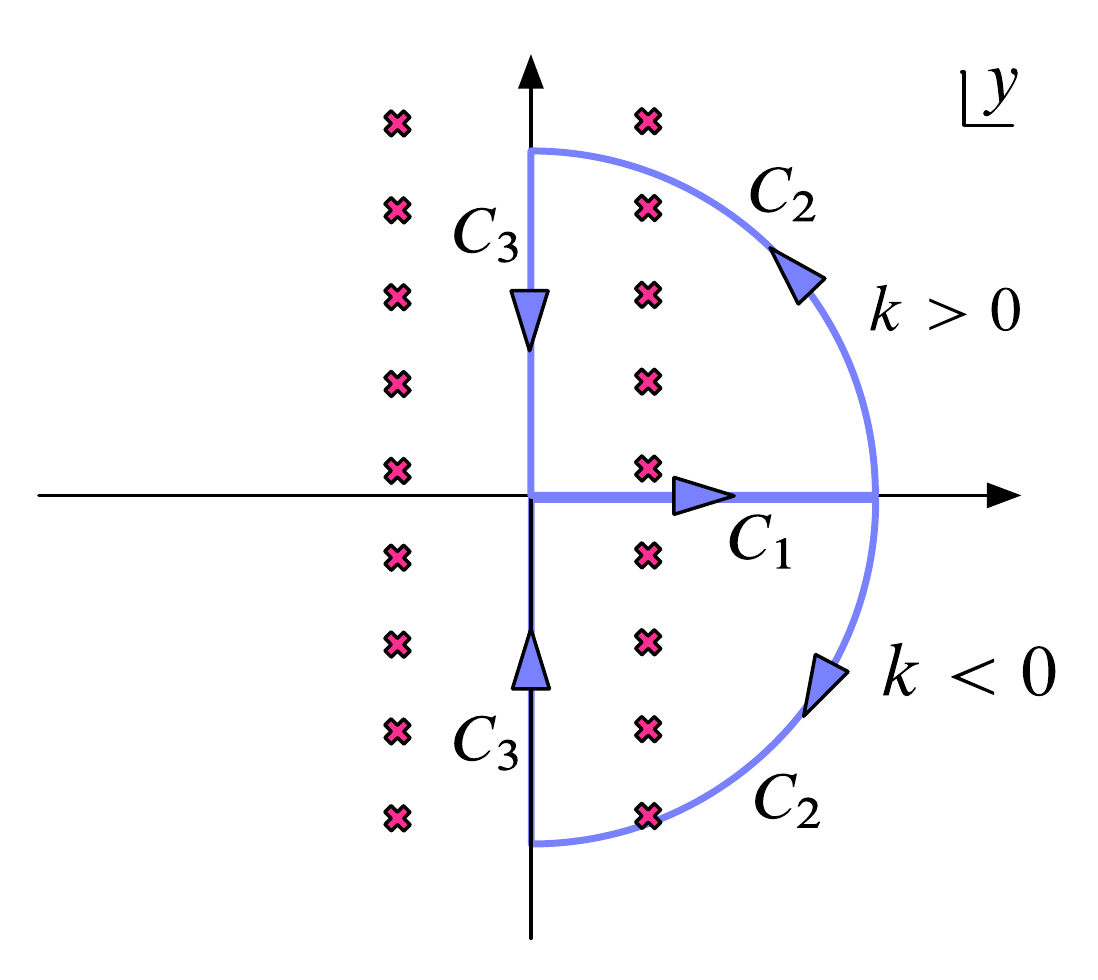}
\caption{The integration contour for $I_2(k)$. The radius of the contour $C_2$ is taken to be~infinite.}
\end{figure}

\end{document}